%% file: Parsai2019STTT.tex
\RequirePackage{fix-cm}

\documentclass[twocolumn]{article}

\usepackage{authblk}
\usepackage{moreverb}
\usepackage{subfig}
\usepackage[normalem]{ulem}
\usepackage{graphicx}
\usepackage{balance}
\usepackage{multirow}
\usepackage{hhline}
\usepackage{adjustbox}
\usepackage{amssymb}
\usepackage{color}
\usepackage{ifthen}
\usepackage[table]{xcolor}
\usepackage[perpage]{footmisc}
\usepackage{paralist} %
\usepackage{xspace}
\usepackage{hyperref}

\usepackage{atbegshi,picture}
\newcommand{\LittleDarwin}{\textsf{{\small LittleDarwin}}\xspace}
\newcommand{\JaCoCo}{\textsf{{\small JaCoCo}}\xspace}
\newcommand{\PItest}{\textsf{{\small PITest}}\xspace}
\newcommand{\Maven}{\textsf{{\small Maven}}\xspace}
\newcommand{\CI}{continuous integration\xspace}
\newcommand{\CIstartofsentence}{Continuous integration\xspace}
\newcommand{\CIcapitals}{Continuous Integration\xspace}

\newcommand{\catSimCov}{\textsf{SimCov}\xspace}
\newcommand{\catLoBHiM}{\textsf{LoB-HiM}\xspace}
\newcommand{\catHiBLoM}{\textsf{HiB-LoM}\xspace}
\newcommand{\catNoB}{\textsf{NoB}\xspace}
\newcommand{\catNoM}{\textsf{NoM}\xspace}

\newcommand{\RQsubdivision}[1]{\vspace{1em}
\noindent
\textbf{#1}~\xspace}

\newcommand{\secref}[1]{Section~\ref{#1}\xspace}
\newcommand{\figref}[1]{Figure~\ref{#1}\xspace} 
\newcommand{\tabref}[1]{Table~\ref{#1}\xspace}

\newcommand{\RQone}{Is it feasible to integrate mutation testing in a \CI system?}
\newcommand{\RQtwo}{Does mutation testing reveal additional weaknesses in the test suite compared to branch coverage?}

\newcommand{\RQthree}{Can we reduce the performance overhead induced by mutation testing to an acceptable level?}

\usepackage{fancybox}
\newcommand{\hypobox}[1]{
\begin{center}%
        \noindent\thicklines\setlength{\fboxsep}{8pt}%
        \cornersize{0.2}\Ovalbox{
        \begin{minipage}{0.85\linewidth}%
        \textit{#1}
        \end{minipage}} 
\end{center}}

\newcommand{\finaltitle}{Comparing Mutation Coverage Against \\Branch Coverage in an Industrial Setting}

\hypersetup{
    colorlinks,%
    citecolor=black,%
    filecolor=black,%
    linkcolor=black,%
    urlcolor=gray,
    linktocpage=true,
    bookmarksopen=true,
    pdfpagemode=UseOutlines,
    pdftitle={\finaltitle},    %
    pdfauthor={Ali Parsai, Serge Demeyer},     %
    pdfsubject={ARXIV VERSION},   %
}

\begin{document}

\title{\finaltitle}
\author{Ali Parsai}
\author{Serge Demeyer}
\affil{ali.parsai@flandersmake.be, serge.demeyer@uantwerpen.be\\ Flanders Make and Universiteit Antwerpen}

\date{}

\maketitle

\begin{abstract}
The state-of-the-practice in software development is driven by constant change fueled by continuous integration servers.
Such constant change demands for frequent and fully automated tests capable to detect faults immediately upon project build.
As the fault detection capability of the test suite becomes so important, modern software development teams continuously monitor the quality of the test suite as well.
However, it appears that the state-of-the-practice is reluctant to adopt strong coverage metrics (namely mutation coverage), instead relying on weaker kinds of coverage (namely branch coverage).
In this paper, we investigate three reasons that prohibit the adoption of mutation coverage in a continuous integration setting: (1) the difficulty of its integration into the build system, (2) the perception that branch coverage is ``good enough'', and (3) the performance overhead during the build. Our investigation is based on a case study involving four open source systems and one industrial system. We demonstrate that mutation coverage reveals additional weaknesses in the test suite compared to branch coverage, and that it is able to do so with an acceptable performance overhead during project build.

\end{abstract}

\maketitle

\input{01-Introduction}

\input{02-Background}

\input{03-Tools}

\input{04-CaseStudyDesign}
\input{05-AnalysisOfResults}

\input{06-ThreatsToValidity}
\input{07-RelatedWorks}

\input{69-Conclusions}

\section*{Acknowledgements}
{\footnotesize We would like to express our gratitude to the  HE/Imaging IT Clinical Applications team at Agfa Healthcare -- Belgium, for allowing us to conduct these analyses on the segmentation component of the Impax ES medical imaging software. This work is sponsored by:\\ (a) ITEA$^3$ \textsf{TESTOMAT Project} (number 16032), sponsored by VINNOVA -- Sweden's innovation agency;\\
(b) Flanders Make vzw, the strategic research centre for manufacturing industry.}

\bibliographystyle{plain}
\balance
\bibliography{Parsai2019STTT}

\end{document}

%% file: 01-Introduction.tex
\section{Introduction}
\label{section:Introduction}
With the popularity of agile methods, especially test-driven development~\cite{Beck2002} and continuous integration~\cite{Fowler2006}, developers shifted their focus to test the changes to their software early and often~\cite{McGregor2007}. This strategy demands a high standard of quality for test suite to avoid potential faults as much as possible in the later stages of software development~\cite{Bjerke-Gulstuen2015}.
Developers assess the quality of test suite according to its capability of detecting yet unknown faults. 
For a faulty statement to be revealed, the program must pass through four stages:
(i) the faulty statement must be executed \textit{(Reachability)}, (ii) the faulty statement needs to affect the program state \textit{(Infection)}, (iii) the effect of the faulty statement on program state need to propagate to the program output \textit{(Propagation)}, and the test needs to observe the failure in the program output \textit{(Reveal)}~\cite{Ammann2016,Li2017}.
Several coverage metrics exist that aim to quantify the quality of a test suite. 
Among them, mutation coverage is generally acknowledged as the state-of-the-art coverage metric since it checks whether a test covers all four aforementioned stages~\cite{Andrews2005,Andrews2006,Harman2011,Just2014a,Papadakis2018}. %
Indeed, mutation testing is the process of deliberately injecting faults into a software system, and then verifying whether the tests actually fail. This faulty version of the software is called a \textit{mutant}. The faults injected by %
each mutant are modeled after the common mistakes often made by developers, hence mutation testing provides a repeatable and scientific approach to measure the fault detection capability of a test suite~\cite{Jia2011,Papadakis2018}. Using the results of mutation testing, it is possible to improve test suite quality or prioritize tests~\cite{Baudry2006,Do2006,Smith2009,Smith2009a}. 
Comparative  studies demonstrated that in terms of fault detection, mutation testing is more effective than several other coverage criteria~\cite{Andrews2005,Frankl1993,Li2009}. 
If the fault model used for mutation operators is close to reality, mutation testing produces more accurate results than simple coverage metrics~\cite{Andrews2006}. 
Finally, mutation testing has been shown to subsume statement and branch coverage~\cite{Offutt1996a}. This is due to the fact that branch and statement coverage only check for reachability~\cite{Harman2011}.

Despite these promising results, the state-of-the-art is not yet adopted into the state-of-the-practice. For instance, Gopinath et al. put forward that mutation testing  \textit{``is generally not used by real-world developers with any frequency"}~\cite{Gopinath2014}. Instead, the state-of-the-practice relies on simple coverage metrics such as statement and branch coverage. This is despite %
their inadequacy for assessing the fault detection capability of a test suite~\cite{Wei2012,Hemmati2015}. Among these metrics, branch coverage is commonly used in industry~\cite{Tengeri2014}. However, even with 100\% of branch coverage there is still the potential for faults to go unnoticed, because  branch coverage only checks whether the test code executes  each branch and does not assess whether the program state is infected or the fault is propagated to the output~\cite{Marick1991,Gopinath2014,Tian2009}. Because of this, such coverage metrics cannot reveal anything regarding the quality of the test oracle.

A possible explanation for the preference towards simple coverage metrics is they are relatively easy and fast to collect. Today, there are plenty of test coverage tools available that instrument the code, execute the tests, and report the parts of the code not covered by the tests (i.e., statements, branches, paths). Also, while the performance overhead of these tools  on the overall test execution  are not negligible, it can often be tolerated in development environments~\cite{Blondeau2017,Chen2018}. Moreover, Gligoric et al. demonstrate that branch coverage---among several coverage criteria---is the best one to predict the mutation coverage of a test suite~\cite{Gligoric2013}. From a practitioner's point of view,  branch coverage  is a reasonable quality criterion  because of the trade-off between time (the performance overhead induced by collecting the measurements) and quality (a ``good enough'' fault detection capability). This means that the successful adoption of mutation testing in industry requires convincing practitioners that there are tangible and worthwhile differences between the two methods when it comes to analyzing industrial software.

Literature blames the lack of adoption of mutation testing in industry mainly on the performance overhead, leading to the adagium ---\textit{do fewer, do smarter, and do faster}~\cite{Offutt2001}. Examples in that sense are mutant sampling, weak mutation testing, and performing the mutation testing on byte code rather than the source code~\cite{Jia2011,Papadakis2018}.
However, the performance overhead is only one part of the equation. We argue that there are two other issues when adopting mutation testing in an industrial strength \CI setting: (i) the complexity of continuous build environments and (ii) the optimism regarding the added value of mutation testing. We expand on each of those issues below.
\begin{compactitem}
\item [(i)] First, the research on mutation testing has ignored the novel trend that modern test infrastructure is part of a \CI environment. In such a setting, the build steps follow one another in lockstep.
In fact, fully automatic test infrastructure is now common place in many companies, e.g. Google performs roughly 800 thousand builds and 150 million test runs in an average day automatically~\cite{Memon2017}.
Integrating additional steps (such as calculating mutation coverage) easily interferes with such automated systems, and requires up-front considerations in the design to ease the integration. %
\item  [(ii)] Second, there is little empirical evidence that mutation testing reveals additional weaknesses in industrial strength test suites. Typically, studies that promote mutation testing are based on open source cases with components designed to be open and accessible~\cite{Frankl1997,Inozemtseva2014,Gligoric2013,Li2009}. The same is not true for the industrial systems, where brown field development is common and legacy code is integrated as black-box components and tested accordingly~\cite{Hopkins2008,Garousi2018}.
\end{compactitem}

\noindent
We derived these issues from a pilot study, integrating mutation testing in an industrial strength \CI setting. In particular, we were asked to explore the advantages and drawbacks of mutation testing in the context of the Segmentation component of the Impax ES medical imaging software used by Agfa HealthCare. The Segmentation component provides imaging algorithms to perform segmentation on 3D modeled volumes. As can be expected from a software system in healthcare, it must adhere to strict safety standards including advanced monitoring of test quality. Yet, the segmentation component interfaced with some legacy code, hence black-box testing was an inherent part of the testing strategy. Also, the Impax ES system is implemented as a small product-line, hence the \Maven build system was configured to resolve dependencies with libraries depending on the product line variant to be built.

In our work, we first explore the feasibility of mutation testing in an industrial project which relies on a \CI system (RQ1).
Then, we investigate the pros and cons that arise by its adoption. We attempt to verify that compared to branch coverage,  
mutation testing has the pros of revealing additional weaknesses in the test suite  (RQ2), and it has the cons of introducing overhead (RQ3). 
This leads us to pursue the following research questions: 

\noindent
\textbf{RQ1:} \emph{\RQone}
  \begin{compactitem}
  \item[$\Rightarrow$] To answer this question we follow a proof by construction. We first integrate an existing mutation coverage tool (namely \PItest) into the build system (namely \Maven) %
We report the challenges we encountered and the workarounds we performed, all to no avail. Consequently, we adapted and used a mutation testing tool named \emph{\LittleDarwin} specifically designed to integrate well within a \CI environment.
  \end{compactitem}

\noindent
\textbf{RQ2:} \emph{\RQtwo}
  \begin{compactitem}
  \item[$\Rightarrow$] To answer this question we analyze branch coverage along with the mutation coverage. 
  We investigate parts of the code where these two metrics differ to see whether mutation coverage indeed exposes additional weaknesses. To increase the generalizability of our findings, we perform the same comparison on four open source systems.
  \end{compactitem}

\noindent
\textbf{RQ3:} \emph{\RQthree} 
  \begin{compactitem}
  \item[$\Rightarrow$] To answer this question, we measure the performance overhead induced by a full mutation analysis, injecting 12K mutants for 38K lines of code. We compare this against the performance overhead after \emph{mutant sampling} (Section \ref{section:mutationtesting}), effectively reducing the number of mutants to 34.7\%. %
   We verify the results of the full mutation coverage against the sampled mutation coverage to see whether the reduced number of mutants still reveals weaknesses in the test suite.
  \end{compactitem}

\noindent						
The rest of paper is structured as follows. In \secref{section:Background}, we provide some background information on test coverage in general and mutation testing in particular. We describe  the main tools used in this study in \secref{section:tools}. We then proceed with a discussion of the case study design, including a description of the cases under investigation as well as a detailed explanation of the setup for the open source and industrial cases in \secref{section:casestudydesign}. The results of our case study are then discussed in \secref{section:analysis}, followed by a discussion of the threats to the validity of this study in \secref{section:threats}. An overview of related work is presented in \secref{section:relatedwork}. Finally, in \secref{section:Conclusions}, we present our final conclusions.

%% file: 02-Background.tex
\section{Background}
\label{section:Background}
In this section we present an overview of the background information necessary to understand the rest of the paper. We briefly introduce \CI environments and discuss the importance of test suite quality therein (\secref{section:ci}), give a brief explanation of code coverage in general and branch coverage in particular (\secref{section:codecoverage}), and discuss mutation testing and its related concepts in detail (\secref{section:mutationtesting}).

\subsection{Testing in \CIcapitals Environment}
\label{section:ci}

\CIstartofsentence is defined as the practice of merging the developed code with a central source code repository as often as possible. The concept of \CI was first proposed by Booch as a way to avoid integration problems~\cite{Booch1991}. This method has been in the center of attention in the past decade since it is the basis for today's agile development techniques. 
\CIstartofsentence allows the software to be continuously tested. Simple tasks (e.g. unit tests) can be triggered upon each commit; 
whereas, the time-consuming tasks (e.g. integration test) can be postponed to the nightly build. Providing continuous feedback, the continuous integration environment takes care that the code-base remains stable during development, and reduces the risk of arriving in integration hell (the point in production when members on a delivery team integrate their individual code)~\cite{Fowler2006}.

The introduction of agile development techniques has resulted in an increased interest in the fault detection capability of the test suite. This is typically monitored by means of \emph{code coverage metrics}.
In this context,  a weakness in a test suite is a lack of capacity of a test suite to detect faults in a particular part of software, either by a lack of tests to cover that part, or incomplete testing of the covered part. 

\subsection{Code Coverage Metrics} 
\label{section:codecoverage}
Code coverage is defined as the proportion of code that is tested by the test suite. There are several ways to calculate code coverage. The most often used metrics in industry are statement coverage and branch coverage~\cite{Gopinath2014}. Statement coverage is the number of statements in the program that are executed at least once by the test suite divided by the total number of statements. Similarly, branch coverage is the number of branches executed at least once by the test suite divided by the total number of branches (Equation~\ref{branchcoverageequation}). Branch coverage subsumes statement coverage, because if all branches are examined, all statements contained in the branches are examined in the process~\cite{Zhu1997,Ammann2016}. 
Branch coverage is often used in popular industrial tools to evaluate the quality of a test suite. More specifically, a high value of branch coverage is assumed to imply a ``good enough'' test suite~\cite{Yang2007}.

\begin{equation}
\label{branchcoverageequation}
\resizebox{0.9\linewidth}{!}{$
Branch\ Coverage = \frac{Number\ of\ branches\ executed\ at\ least\ once}{Number\ of\ all\ branches}
$}
\end{equation}

A test suite that achieves 100\% coverage according to a certain coverage criterion is called an \textit{adequate test suite}. For example, a test suite is branch-adequate when its branch coverage is 100\%. However, it is known that statement-adequate or branch-adequate test suites are ineffective for assessing the fault detection capability of a test suite~\cite{Hutchins1994,Wei2012,Inozemtseva2014,Hemmati2015}. %
An alternative test coverage metric that is used mostly in safety-critical context is decision coverage, namely the proportion of decision points triggered by tests. The avionics standard RTCA/DO-178C and the automotive standard ISO26262  enforce complete modified condition/decision coverage (MC/DC)~\cite{Brosgol2011,ISO2011}. Writing tests that achieve complete MC/DC is very difficult, and needs a high level of expertise in the system under test~\cite{Hayhurst2001}. Yet, even 100\% MC/DC does not guarantee the absence of faults~\cite{Gay2015,Kandl2014}.

\subsection{Mutation Testing}
\label{section:mutationtesting}

Mutation testing is the process of injecting faults into a software system and then verifying whether the test suite indeed fails (i.e. detects the injected fault). The idea of mutation testing was first mentioned in a class paper by Lipton (as reported by Offutt et al. in~\cite{Offutt2001}), and later developed by DeMillo, Lipton and Sayward~\cite{DeMillo1978}. The first implementation of a mutation testing tool was done by Timothy Budd in 1980~\cite{Budd1980}.

Mutation testing induces the following steps on the test process.  It starts with a \textit{green} test suite --- a test suite in which all the tests pass. First, a faulty version of the software is created by introducing faults into the system \textit{(Mutation)}. This is done by applying a known transformation \textit{(Mutation Operator)} on a certain part of the code.  After generating the faulty version of the software \textit{(Mutant)}, it is passed on to the test suite. If there is an error or failure during the execution of the test suite the mutant is marked as killed \textit{(Killed Mutant)}. If all tests pass, it means that the test suite could not catch the fault and the mutant has survived \textit{(Survived Mutant)}.

\subsubsection*{Invalid Mutants.}
In the process of generation of mutants, sometimes a mutant is not compilable. Such mutants are called \emph{invalid mutants}. Given the fact that typical mutation testing tools do not attempt to compile the code entirely, it is possible that mutants  are created that adhere to the syntax of a language, but cannot be compiled. For example, in case of concatenation of two string variables using ``+'' operator, changing this operator to ``-'' leads to generation of an invalid mutant. While most invalid mutants can be avoided at mutant generation time, some are difficult to filter out without having the facilities of a compiler.  %

\subsubsection*{Equivalent Mutants.}
\begin{figure}
\centering
		\includegraphics[width=\linewidth]{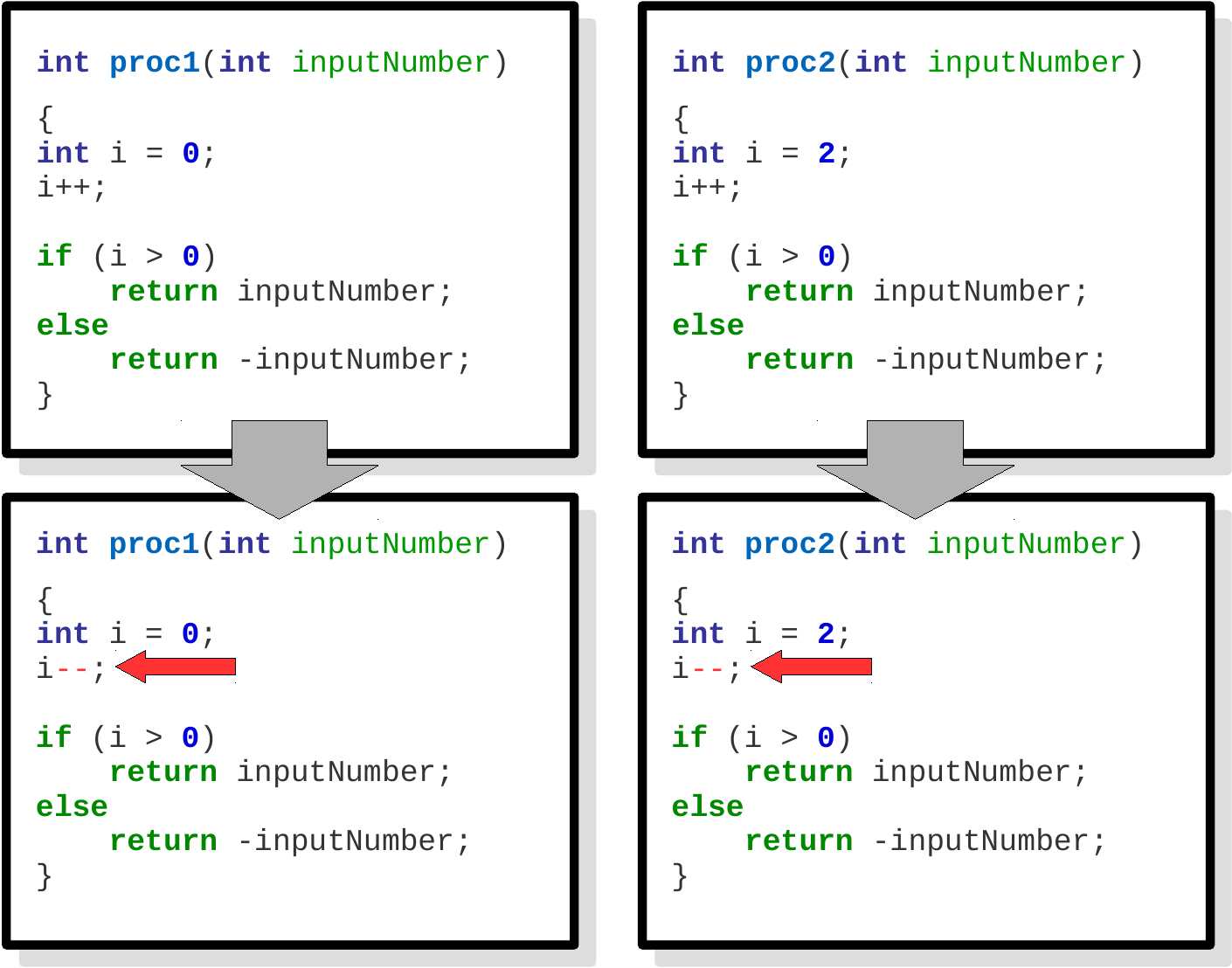}
		\caption{Example of an equivalent mutant in \texttt{proc2}}
		\label{fig:equivalent-context}
\end{figure}

If the output of a mutant for all possible inputs is the same as the original program, it is called an \emph{equivalent mutant}. It is not possible to create a test case that passes for the original program and fails for an equivalent mutant, because the equivalent mutant has the same semantics as the original program. This makes the  creation of equivalent mutants undesirable, since the time that the developer wastes on an equivalent mutant does not result in the improvement of the test suite. Equivalent mutants have a significant impact on the accuracy of the mutation coverage~\cite{Grun2009}. Unfortunately, equivalent mutants are not easy to detect because they depend on the context of the program itself~\cite{Madeyski2014}. For example, in \figref{fig:equivalent-context}, $--$ replacing $++$ in \textit{proc1}  changes the output for any input other than $0$, while the same mutant in \texttt{proc2} does not. Indeed, the preceding line $i++$  ensures that the condition $i > 0$ is always met for $i \geq 1$. The mutant can be killed in  \texttt{proc1}, because $i = 0$, however in \texttt{proc2} the mutant is undetectable by any test because of $i = 2$.

As for filtering the equivalent mutants, there are no tools available that automatically detect and remove all equivalent mutants. In general, detection of equivalent mutants is an undecidable problem~\cite{Offutt1997}. Manual inspection of all mutants is the only way of filtering all equivalent mutants, which is impractical due to the amount of work it needs. Therefore, the common practice within today's state-of-the-art is to take precautions to remove as many equivalent mutants as possible (e.g. using Trivial Compiler Equivalence~\cite{Papadakis2015}), and accept equivalent mutants as a threat to validity.

\subsubsection*{Mutation Coverage.} 
Mutation testing allows software engineers to monitor the fault detection capability of a test suite by means of \emph{Mutation Coverage} (see Equation~\ref{coverageequation}).
A test suite is said to achieve \emph{full mutation test adequacy} whenever it can kill all of the non-equivalent mutants, thus reaches a mutation coverage of 100\%. Such test suites are called  \emph{mutation-adequate test suites}. 

\begin{equation}
\label{coverageequation}
\resizebox{0.9\linewidth}{!}{$Mutation\ Coverage = \frac{Number\ of\ killed\ mutants}{Number\ of\ all\ non\mbox{-}equivalent\ mutants}$}
\end{equation}

Mutation coverage is often declared as a \emph{stopping criterion} for writing (unit) tests --- the next level of testing can only start when mutation coverage exceeds a given threshold~\cite{Offutt1996b,Zhu1997}. %
This is especially useful when tests are generated automatically~\cite{DeMilli1991,Fraser2012}. %

\subsubsection*{Mutation Operators.}

A mutation operator is a known transformation which creates a faulty version by introducing a single change. The first set of the mutation operators designed were reported in King et al.~\cite{King1991}. These operators which work on very basic entities were introduced in the tool Mothra which was designed to mutate FORTRAN77 programming language. In 1996, Offutt et al. determined that a selection of few mutation operators are enough to produce similarly capable test suites  with a four-fold reduction of the number of mutants~\cite{Offutt1996}. This reduced set of operators  shown in \tabref{table:reducedsetofoperators} remained more or less intact in all subsequent research papers. 

\begin{table}[!h]
	\footnotesize
	\centering
	\caption{Reduced-set mutation operators (adapted from~\cite{Ma2006a} \textcopyright ACM 2006)}
	\label{table:reducedsetofoperators}
	\begin{tabular}{|c||c|}
		\hline \textbf{Operator} & \textbf{Description} \\ 
		\hline \hline AOR & Arithmetic Operator Replacement \\ 
		\hline AOD & Arithmetic Operator Deletion \\ 
		\hline AOI & Arithmetic Operator Insertion \\ 
		\hline ROR & Relational Operator Replacement \\ 
		\hline COR & Conditional Operator Replacement \\ 
		\hline COD & Conditional Operator Deletion \\ 
		\hline COI & Conditional Operator Insertion \\ 
		\hline SOR & Shift Operator Replacement \\ 
		\hline LOR & Logical Operator Replacement \\ 
		\hline LOD & Logical Operator Deletion \\ 
		\hline LOI & Logical Operator Insertion \\ 
		\hline ASR & Assignment Operator Replacement \\ 
		\hline 
	\end{tabular} 
	
\end{table}

With the popularity of the object-oriented programming paradigm, there was a need to design new mutation operators to simulate the faults that occur  in this kind of programs. Several studies proposed new mutation operators~\cite{Ma2002,Chen2006}, and some of them were designed to prove the usefulness of object-oriented operators~\cite{Lee2004,Derezinska2012}. Ahmed et al.  did a complete survey on this subject~\cite{Ahmed2010}.

During the past decade, the academic focus was on creating new mutation operators for special purposes such as targeting certain security problems~\cite{Shahriar2008a,Zeng2009} or language specific mutation operators~\cite{Abraham2009,Bradbury2006a,Silva2012,Parsai2018,Parsai2019}. %
These mutation operators, even though important in their own context, do not relegate into the general concept of mutation testing. The traditional mutation operators are by far the most often implemented~\cite{Papadakis2018}. One reason for this is that using more mutation operators produces more mutants; which makes the procedure  longer to finish, and as a result, less practical. The reduced set of operators mentioned in \tabref{table:reducedsetofoperators} provides a smaller set which produces results with enough detail for any practical purpose, even though the confidence in such results are slightly less than those retrieved by using additional mutation operators.

\subsubsection*{Mutant Sampling.}
\label{section:sampling}
To make mutation testing practically applicable, it is important to reduce the time needed --- do fewer, do smarter, and do faster~\cite{Offutt2001}. ``Do fewer'' is achieved by \emph{mutant sampling}: randomly selecting a sample set of mutants instead of processing all of them. This idea was first proposed by Acree~\cite{Acree1980} and Budd~\cite{Budd1980} in their PhD theses. Since then, there were many studies confirming the effectiveness of this approach: the performance gain is significant yet reveals the same weaknesses~\cite{King1991,Wong1993,Mathur1994,Zhang2010}. The random mutant selection can be performed uniformly, meaning that each mutant has the same chance of being selected. Otherwise, the random mutant selection can be enhanced  by using heuristics based on  the source code.

The percentage of mutants that are selected determines the \textit{sampling rate} for random mutant selection. Using a fixed sampling rate is common in literature~\cite{Wong1995,Zhang2010,Zhang2013}. However, it is possible to use a weight factor to optimize the sampling rate according to various parameters such as the number of mutants per class. This is called \textit{weighted} mutant sampling~\cite{Parsai2016}. It is also possible to determine the sampling rate dynamically while performing mutation testing. A method resembling the latter was proposed by Sahinoglu and Spafford to randomly select the mutants until the sample size becomes statistically appropriate~\cite{Sahinoglu1990}. They concluded that their model achieves better results due to its self-adjusting nature~\cite{Jia2011}. 

There is one other factor besides the sampling rate that needs to be considered when sampling; the total amount of time that is practically viable. Unfortunately, in the current literature we did not find any concrete targets. Therefore, we set our own target based on a hypothetical scenario of an agile team running the whole mutation testing once every week during the weekend. In this scenario, the team works from Monday at 8am till Friday at 6pm, which leaves the whole weekend (thus 62 hours) to perform the analysis.

%% file: 03-Tools.tex
\section{Tools Used in This Study}
\label{section:tools}

In this section we present the test coverage tools used to investigate the trade-offs between branch coverage and mutation coverage in an industrial setting. These tools are \JaCoCo, a tool to compute statement and branch coverage for Java software systems (\secref{section:jacoco}); \PItest a state-of-the-art mutation testing tool designed for easy integration with current test and build tools (\secref{section:pitest}) and \LittleDarwin, the tool we adapted and used to perform mutation testing on Java software systems with complicated build systems (\secref{section:littledarwin}). %
  
\subsection{\JaCoCo.}
\label{section:jacoco}

\JaCoCo [\url{http://www.eclemma.org/jacoco}] is a lightweight, flexible, and well documented tool to provide statement and branch coverage for Java programs. \JaCoCo is compatible with most Java build systems, hence is easy to deploy in a \CI environment.
\JaCoCo is the de facto standard for measuring test coverage for Java projects, and is used as baseline in research concerning test coverage (e.g. ~\cite{Li2013,Janjic2013,Bauersfeld2014,Kracht2014}). %

\JaCoCo uses a set of different probes %
 to calculate coverage metrics. All these probes %
 are instrumented into %
Java class files which  are Java byte code instructions and debug information optionally embedded therein. Consequently, \JaCoCo uses dynamic analysis to compute coverage over byte code which allows it to work even without the source code available. The link to source files are then generated using the debug information that accompanies Java byte code.

However, byte code instrumentation has known disadvantages~\cite{Li2013,Tengeri2016}: By performing a study on \JaCoCo, Tengeri et al. identify 6 reasons why using byte code instrumentation in Java language might not be as accurate as source code instrumentation~\cite{Tengeri2016}. In particular, (i) the act of instrumentation itself can affect the behavior of the tests, (ii) cross-coverage among submodules is not taken into account, (iii) untested submodules are excluded from the analysis, (iv) method signatures are different in byte code and source code for methods that contain compiler-injected parameters, (v) exceptions leading to interruption of the control flow result in loss of information, and (vi) generated code produces obstacles in collecting coverage information. Consequently, the results from \JaCoCo---especially when statements or branches are reported as \emph{not} being covered by a test---need to be double-checked for accuracy.

\JaCoCo is used throughout this study to compute branch coverage for industrial and open source cases.

\subsection{\PItest}
\label{section:pitest}

\PItest [\url{http://pitest.org/}] is a state-of-the-art mutation testing system for Java, designed to be fast and scalable. 
\PItest seamlessly integrates with today's test and build tools (i.e. \textsf{\small Ant}, \textsf{\small Gradle} and \Maven).  \PItest is the de facto standard for mutation testing within Java, %
and it is used as baseline in research concerning mutation testing (e.g.~\cite{Kaczanowski2012,Inozemtseva2014,Parsai2014,Saleh2014,Assylbekov2013}).

\PItest has a wide range of mutation operators. The default setting is practically a subset of the reduced-set of mutation operators (\tabref{pitestmutationoperators}), however because they are applied to byte code, they are grouped and named differently. There are several other mutation operators that %
can be enabled as well. %

\PItest comes with a lot of internal optimizations to tackle the ``do faster'' part of the maxim: \textit{do fewer, do smarter, and do faster}~\cite{Offutt2001}. Most importantly, mutations are performed at the level of byte code to avoid recompilation. 
However, byte code level mutation presents other obstacles to overcome. For example, \PItest needs to find and execute tests by itself for each mutant, which causes incompatibility with complicated build structures. This is more apparent when test code is located in separate packages, and therefore, not easily discoverable by \PItest.
In addition, \PItest incorporates some heuristics to choose which tests to run, which implies that the accuracy of these heuristics affects the results of \PItest as well. In particular, \PItest uses the same mechanism as \JaCoCo to determine statement coverage, and skip the evaluation of mutants in uncovered statements. This, in turn, raises similar issues as described in \secref{section:jacoco}.

\PItest is used in this study in an attempt to demonstrate the feasibility of mutation testing in an industrial environment (RQ1).

\begin{table}[!h]
	\centering
	\caption{\PItest mutation operators at the time of this study} %
	\label{pitestmutationoperators}
\adjustbox{max width=\linewidth}{	\begin{tabular}{|l||l|c|c|}
		\hline \multirow{2}{*}{\textbf{Operator}} & \multirow{2}{*}{\textbf{Description}} & \multicolumn{2}{c|}{\textbf{Example}} \\
		\hhline{~~--} & & \textbf{Before} & \textbf{After} \\ 
		\hline
		\hline CBM & Mutates the boundry conditions  & $a > b$  & $a >= b$ \\ 
		\hline IM & Mutates increment operators & $a++$ & $a--$ \\ 
		\hline INM & Inverts negation operator & $-a$ & $a$ \\ 
		\hline MM & Mutates arithmetic \& logical operators & $a\,\&\,b$ & $a\,|\,b$ \\ 
		\hline NCM & Negates a conditional operator & $a == b$ & $a != b$ \\ 
		\hline RVM & Mutates the return value of a function & return true; & return false; \\ 
		\hline VMCM & Removes a void method call & voidCall(x); & --- \\ 
		\hline 
	\end{tabular}}
	
\end{table}

\subsection{\LittleDarwin.}
\label{section:littledarwin}

\LittleDarwin [\url{http://littledarwin.parsai.net/}] is a mutation testing tool created by Ali Parsai (first author of this paper) to provide mutation testing within a \CI environment.
It is designed to have a loose coupling with the test infrastructure, instead relying on the build system to run the test suite. \LittleDarwin imposes two restrictions only: (a) the build system must be able to run the test suite; (b) the build system must return non-zero if any tests fail, and zero if it succeeds. For a detailed description of \LittleDarwin, please refer to Parsai et al.~\cite{Parsai2017}.
 
For the purposes of this study, there are 9 mutation operators implemented in \LittleDarwin listed \tabref{mutationoperators}. These operators are a subset of the reduced-set of mutation operators (\tabref{table:reducedsetofoperators}), and similar to the default setting of \PItest, but differently grouped and named. %
Since the number of mutation operators of \LittleDarwin is limited, it is possible that no mutants are generated for a class. In practice, we observed that usually only very small compilation units (e.g. interfaces, and abstract classes) are subject to this condition.  
 
 \begin{table}[!h]
 	\centering
 \caption{\LittleDarwin mutation operators}
\label{mutationoperators}
\adjustbox{max width=\linewidth}{
 	\begin{tabular}{|l||l|c|c|}
 		\hline \multirow{2}{*}{\textbf{Operator}} & \multirow{2}{*}{\textbf{Description}} & \multicolumn{2}{c|}{\textbf{Example}} \\
 		\hhline{~~--} & & \textbf{Before} & \textbf{After} \\ 
 		\hline
 		\hline AOR-B & Replaces a binary arithmetic operator & $a + b$  & $a - b$ \\ 
 		\hline AOR-S & Replaces a shortcut arithmetic operator & $++a$ & $--a$ \\ 
 		\hline AOR-U & Replaces a unary arithmetic operator & $-a$ & $+a$ \\ 
 		\hline LOR & Replaces a logical operator & $a\,\&\,b$ & $a\,|\,b$ \\ 
 		\hline SOR & Replaces a shift operator & $a >> b$ & $a << b$ \\ 
 		\hline ROR & Replaces a relational operator & $a >= b$ & $a < b$ \\ 
 		\hline COR & Replaces a binary conditional operator & $a\:\&\&\:b$ & $a\,||\,b$ \\ 
 		\hline COD & Removes a unary conditional operator & $!\,a$  & $a$ \\ 
 		\hline SAOR & Replaces a shortcut assignment operator & $a\:*= b$ & $a\:/= b$ \\ 
 		\hline 
 	\end{tabular}
 }
 \end{table}

For the moment, \LittleDarwin is not optimized for speed. For each mutant injected, \LittleDarwin demands a complete rebuild and test cycle on the build system. This easily leads to several hours of analysis time. Currently the only way to speed-up \LittleDarwin is to use mutant sampling, which is covered under RQ3. 

\LittleDarwin is used throughout this study to compute mutation coverage for industrial and open source cases.

%% file: 04-CaseStudyDesign.tex
\section{Case Study Design}
\label{section:casestudydesign}
In this section, we explain the details about the setup of our case study. We start describing the industrial case (\secref{section:industrialcase}) and open source cases (\secref{section:opensourcecase}). Then, we report the setup of the tools (in \secref{section:toolsetup}). Finally, we provide an overview of the comparison criteria  (in \secref{section:comparisoncriteria}). 

\subsection{Industrial Case}
\label{section:industrialcase}
There are three main components which create the core of Impax ES Clinical Applications. One of these three components is the Segmentation component. The main use of this component is to provide imaging algorithms to segment 3D volumes. This component is average in size compared to the other components of the system, and it includes a test suite which was %
under active development at the time of the case study. The component is entirely  written in Java. %

The team is geographically dispersed across four cities around the globe, thus making coordination a critical part of the software development. This also increases the importance of the test suite during build, because faults discovered downstream require long distance communication over different time zones. The team uses the SCRUM model of development, by holding a weekly sprint meeting to coordinate their efforts~\cite{Rising2000}. %
They use a typical \CI with separate build servers and source code repository servers, centered around  \Maven. The build configuration includes several plugins to compute code coverage, generate reports, and obfuscate the target classes. 

\begin{figure}
	\centering
	\includegraphics[width=\linewidth]{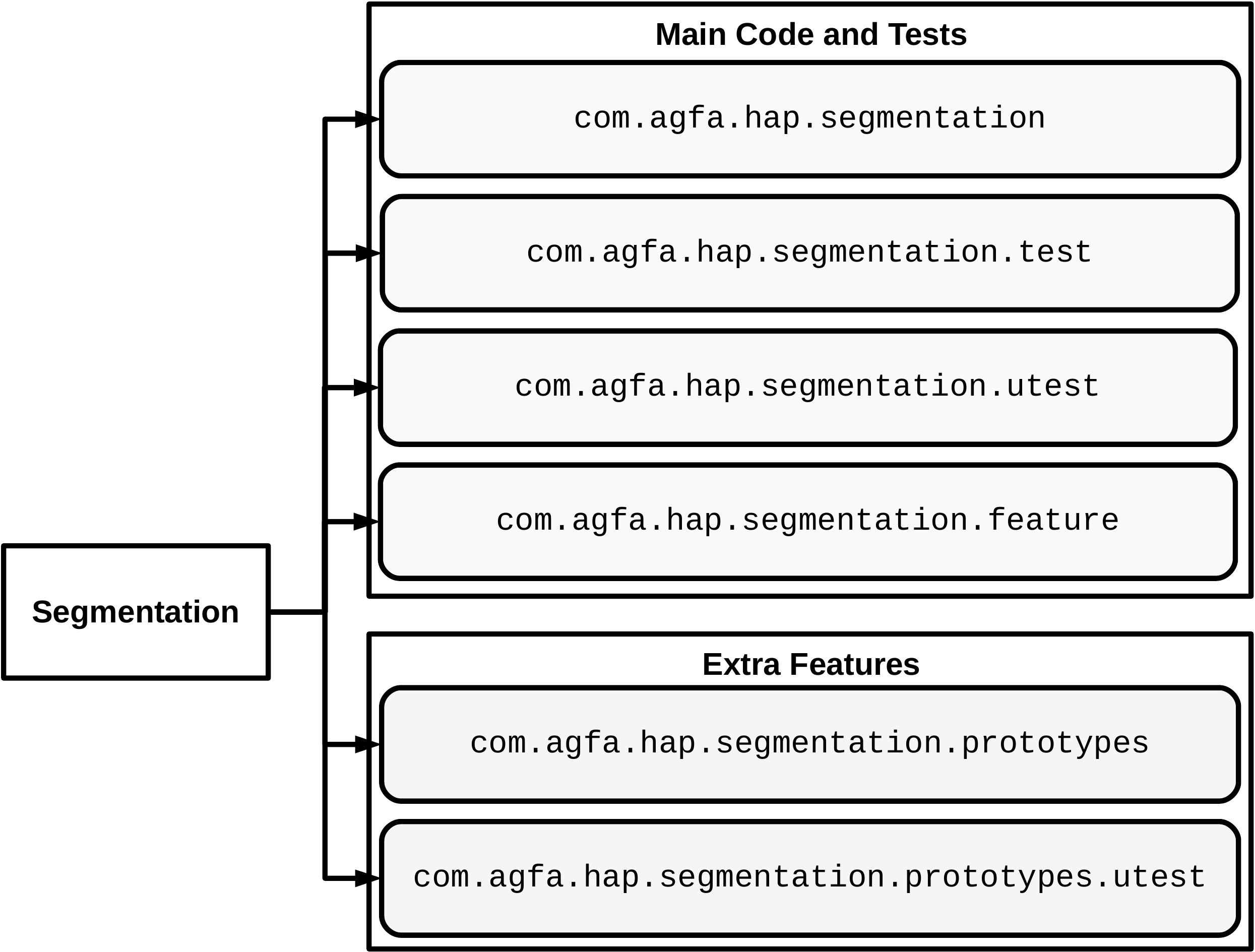}
	\caption{Agfa HealthCare Segmentation build components}
	\label{fig:segmentationbuild}
\end{figure}

The Impax ES system is released in two main variants (production or prototype). There are a few minor variants as well, depending on the target hardware platform the system is deployed upon. Therefore, the Maven build system was configured to resolve dependencies with libraries depending on the product line variant to be built. The system architecture itself relies on the OSGI (Open Service Gateway Initiative) dynamic component model, to load and unload components dynamically without rebooting the system. The extensive use of (dynamic) OSGI headers implies a complicated build process where the \Maven plug-in Tycho [\url{https://eclipse.org/tycho/}] is used to fetch dependencies, compile source files, and run the test suite.

\noindent
The Segmentation component itself is divided into 6 subcomponents as shown in \figref{fig:segmentationbuild}. The main code is located in \textit{com.\-ag\-fa.\-hap.\-seg\-men\-ta\-tion} while the test suite code, some resources used by unit tests and a feature manifest for eclipse are located in the \textit{com.\-ag\-fa.\-hap.\-seg\-men\-ta\-tion.\{utest, test, feature\}} subcomponents respectively. The other two subcomponents provide early prototype features and their unit tests which should be excluded from the final build. Each subcomponent has a separate \textit{pom.xml} file and can be built on its own. There is also a parent \textit{pom.xml} file which selects and builds these subcomponents based on the profile of the product-line variant to be built (production or prototype).

For our case study, we focus on unit test suite of Segmentation component, %
and we do not include the acceptance tests. Removing the acceptance tests decreases the total time for compilation and testing of the Segmentation component from a few hours to less than a minute in each build. %

\subsection{Cases Under Investigation}
\label{section:opensourcecase}

To increase the generalizability of our findings we analyzed four open source systems for addressing RQ2. 
The descriptive statistics of all cases under investigation (the industrial one + the four open source ones) are reported in \tabref{table:casestatistics}. %
Because the non-disclosure agreement does not allow us to reveal too many details, we only list approximate statistics for the industrial case.

\begin{table*}[!h]
\footnotesize
\centering
	\caption{Descriptive statistics for cases under investigation}
\label{table:casestatistics}

\begin{tabular}{| p{4cm} | p{5cm} | p{1cm} |c|c|c|}
\hline \multirow{2}{*}{\textbf{Case}} & \multirow{2}{*}{\textbf{URL}} & \multirow{2}{*}{\textbf{Version}} & \multicolumn{2}{c|}{\textbf{Size (LoC)}} & \multirow{2}{*}{\textbf{Ratio}} \\
\hhline{~~~--~} & & & \textbf{Main} & \textbf{Test} & \\ 
\hline \multicolumn{6}{|l|}{\textit{Industrial Case} {\footnotesize (The numbers are approximated for confidentiality reasons.)}}  \\
\hline Agfa Segmentation & \url{http://www.agfahealthcare.com/global/en/main/resources/product_images/impax_6_0.jsp} & 3.7-snapshot  & 38K & 50K & $\sim$1.3 \\ 
\hline \multicolumn{6}{|l|}{\textit{Open Source Cases}}  \\
\hline Joda Time & \url{http://www.joda.org/joda-time/} & 2.8  & 28479 & 54645 & 1.92 \\ 
\hline Apache Commons Codec & \url{http://commons.apache.org/proper/commons-codec/} & 1.7   & 5773 & 9917 & 1.72 \\ 
\hline jOpt Simple & \url{http://pholser.github.io/jopt-simple/} & 5.0 &  1958 & 6072 & 3.10 \\ 
\hline AddThis Codec & \url{http://github.com/addthis/codec} & 3.2.1 & 3614 & 1318 & 0.36 \\ 
\hline 
\end{tabular} 
\end{table*}

\subsection{Tool Setup}
\label{section:toolsetup}

\JaCoCo was already in use as part of the \Maven build configuration in the industrial case and some of the open source cases, therefore we used \JaCoCo to calculate the branch coverage for the rest as well. For this case, \JaCoCo was being used with its default parameters. %
\PItest was run using the default suite of mutation operators, and it was run in parallel mode, which detects the number of available CPU cores and uses all of them for the analysis.  %
Finally, \LittleDarwin was run with two sets of commands: the first to compile the mutated source code and install the final result into the local \Maven repository; the second to execute the test suite on the compiled version. This was necessary because the production code and the test code have separate build systems.

To make the performance comparison (RQ3) valid, we ran the analysis of all open source cases on the same machine, operating system, and python interpreter. However, since the industrial case was analyzed 5 months prior to the open source cases on premise, the analysis was done on different hardware and operating system, hence the absolute numbers of the execution times cannot be compared. 
For the industrial case, the machine used has two Intel Xeon 2.80 GHz processors with 16 GB (4x4 GB) of DDR3-1333 memory running Windows 7 Enterprise Edition. Since \LittleDarwin only uses a single thread to perform its analysis, there are no gains from the multi-core architecture of the hardware it runs on. %
The open source cases were analyzed on a custom made PC with AMD 1090T 3.2 GHz processor and 8 GB (2x4 GB) memory running Linux Mint 17. 
The execution time for a build is extracted from the output of the build system (which was \Maven in all our cases). The execution time for \JaCoCo was extracted in the same manner, since \JaCoCo acts as a \Maven plugin. The execution time for \LittleDarwin was calculated by aggregating the time spent generating the mutants and the time  spent for gathering the results of the execution of the tests for all mutants. 

\subsection{Comparison Criteria}
\label{section:comparisoncriteria}

In this paper we analyze branch and mutation coverage from a conceptual point of view. For both metrics, a higher value is assumed to suggest a good test suite quality  to the developer. While branch coverage is widely used  in industry, mutation coverage is known to be a better indication of fault detection capability of a test suite. For this reason, we are interested to explore the situation where  branch and mutation coverage present different values.

Considering $m$ as mutation coverage percentage\footnote{Mutation coverage is often expressed as a ratio rather than a percentage, but for the sake of consistency, we use it as a percentage here.}, $b$ as the branch coverage percentage, and $t$ as a threshold, we define 5 categories (\tabref{table:categories}). %
Note that we consider coverage at class level, hence the unit of analysis is a \emph{class}.

\begin{table*}[!h]
	\centering
		\caption{Categorization of differences between branch and mutation coverage}
\label{table:categories}

	\adjustbox{max width=\linewidth}{	
		\begin{tabular}{|l|l|l|}
			\hline Category  & Definition  & Description \\ 
			\hline\hline  \catSimCov & $(|m-b| <= t \land m,b>0) \lor m=b=0$ & Similar branch and mutation Coverage  \\ 
			\hline  \catLoBHiM &  $m-b > t \land m,b>0$   &  Low Branch coverage and High Mutation coverage \\ 
					~ & \multicolumn{2}{r |}{(confirms the fault detection capability of the test suite)} \\ 
			\hline  \catHiBLoM &  $b-m > t \land m,b>0$  &  High Branch coverage and Low Mutation coverage \\ 
					~ & \multicolumn{2}{r |}{(false confidence regarding the fault detection capability of the test suite)} \\ 
			\hline \textsf{NoB} & $m > 0 \land b=0$ &  No Branch coverage \\ 
			\hline  \textsf{NoM} & $b > 0 \land m=0$  &  No Mutation coverage \\ 
			\hline 
		\end{tabular} 
		 }
	\end{table*}

\begin{compactitem}
\item The Category~\catSimCov corresponds to the category in which the difference between $m$ and $b$ is less than a given threshold ($t\%$). For these classes mutation coverage does not provide extra information with respect to branch coverage.
\item For the category~\catLoBHiM the mutation coverage is larger than the branch coverage. There mutation testing provides extra confidence concerning the test suite; despite the low branch coverage the test suite has a high fault detection capability. \emph{This category represents those classes where branch coverage is ``good enough''.}
\item In contrast, the category~\catHiBLoM marks classes where the mutation coverage is smaller than the branch coverage. This is the most interesting category for our investigation. Indeed, mutation testing reveals weaknesses in the test suite, namely where test suite lacks of detectability of a potential fault (Section \ref{section:ci}). From another point view, \emph{this category shows where high branch coverage gives a sense of false confidence regarding the fault detection capability of the test suite.} %
\item Finally, the categories~\catNoB and~\catNoM mark the special cases where the corresponding coverage metric is zero. If both the branch and mutation coverage are zero, it most likely corresponds to a class which is never tested. However, for \catNoB, this may also be due to anomalies in the byte code level instrumentation of \JaCoCo (see \secref{section:jacoco}). When \catNoM is zero this is most likely caused by lack of mutable statements in the code (see \secref{section:littledarwin}). %
\end{compactitem}

\begin{figure*}
	\centering
	\includegraphics[width=0.9\linewidth]{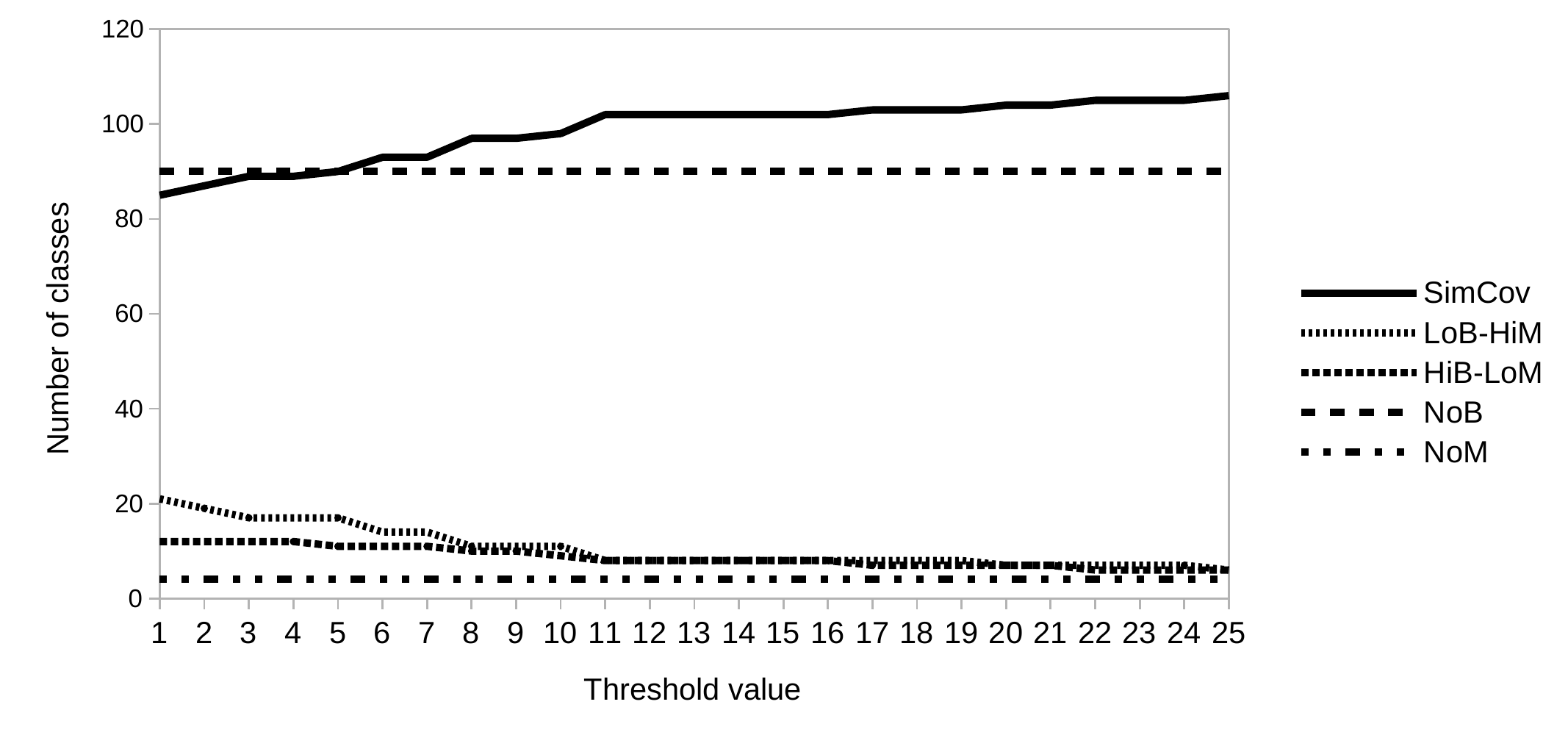}
	\caption{Number of classes in each category for $t$ values between 1 and 25 in the industrial case}
	\label{fig:threshold}
\end{figure*}

The value of $t$ determines the threshold that the two coverage scores are considered close enough so that the difference between them does not make any practical difference. For example, for a threshold of $10\%$, if for a particular class the branch coverage is $55\%$ and mutation coverage is $63\%$, having either of these scores does not change the perceived test quality of that class, and the coverage can be interpreted as ``around $60\%$''. Therefore, by adjusting $t$, we can model the sensitivity of developers to the coverage score.  This threshold does not have any effect on the number of classes in Categories~\catNoB and~\catNoM, and only changes the number of classes in Categories \catSimCov, \catLoBHiM and \catHiBLoM. To determine $t$ we examine the results of both metrics, and choose the minimum value of $t$ where neighboring  $t$ values would not change the number of classes in each category. This way, the threshold is selected based on the specifics of each case, minimizing the analysis bias. For our investigation, we tried all ordinal values of $t$ between 1 and 25 and ultimately derived $t=11$ and $t=7$ respectively for the industrial and the open source cases. The result of this derivation for the industrial case is shown in \figref{fig:threshold}. The number of classes in different categories remain fairly constant after $t=11$, hence is chosen as the threshold. %

Another way to compare mutation and branch coverage is to compute the correlation of these two metrics. Specifically, we analyze whether the order of a set of classes by one metric can predict the order by another set using Kendall's $\tau_b$ coefficient as described in statistic handbooks~\cite{Abdi2007,Agresti2010}. This has been used previously in a similar study by Gligoric et al. where they argue that this coefficient is more appropriate than a linear correlation coefficient because it does not assume a linear correlation between the metrics~\cite{Gligoric2013}.

%% file: 05-AnalysisOfResults.tex
\section{Results and Discussion}
\label{section:analysis}
In this section, we discuss the results of our study. For each research question, we briefly describe our motivation, then our approach, and conclude with our observations. 

\subsection*{\textit{RQ1. \RQone}} 

\RQsubdivision{Motivation.} Mutation testing adds the following steps to the build process: (a) inject a mutant into the software system, (b) build the mutated system, (c) execute the test suite, (d) record the results, (e) restore the system to its original state, and (f) repeat the procedure until there is no more mutants. These steps easily interfere with other parts of the build process (i.e. selection of product line variants, dependency resolution, and dynamic component loading) hence the reason to address the feasibility issue. 

\RQsubdivision{Approach.}
To answer this question we follow a proof by construction. We integrate a state-of-the-art mutation coverage tool (namely \PItest) into the build environment of the industrial case (namely \Maven). The industrial case is representative for many other industrial systems: it has legacy components (where unit tests are missing), it has two major variants (incorporated in a product-line architecture), and has a complicated build structure (where components are dynamically loaded into the build environment by means of OSGI). We report the challenges we encountered and the workarounds we performed, all to no avail. Consequently, we adapted and used a mutation testing tool named \LittleDarwin specifically designed to integrate well within a \CI environment.

\RQsubdivision{Findings.} 
We considered a series of tools during our feasibility study: \PItest, \textsf{\small Cheshire}, and \textsf{\small MuUnit}. \PItest is a widely-used mutation testing tool aimed at industrial projects; Cheshire is a  tool to convert OSGI interfaces and MuUnit is a mutation testing tool designed to perform its analysis on OSGI projects.
 \noindent
The challenges we faced performing the proof by construction are summarized as follows:

\begin{compactitem}
\item \textbf{OSGI.} As explained in \secref{section:industrialcase}, the industrial case heavily relies on the OSGI (Open Service Gateway Initiative) dynamic component model for dynamic loading and unloading of components. We first considered to refactor the system and remove the OSGI headers. However, we learned that these OSGI headers are deeply embedded in all of the source code. Removing the OSGI headers would alter the code beyond recognition hence was not an option. 

\item \textbf{\PItest.} 
\PItest does not refer to OSGI in its documentation, nor did we find any other information sources. We tried it out ourselves, and quickly discovered that \PItest could not run the OSGI-dependent code by itself, and the Tycho plug-in for \Maven was incompatible as well. We posted a few questions in the \PItest forums and there it was confirmed that OSGI could cause problems (see \url{https://groups.google.com/d/topic/pitusers/IH21Q4jJaco/discussion}). %
In particular, there were two blocking issues that we encountered during our attempt. First, \PItest cannot handle a test suite in a completely separated package, loaded dynamically via OSGI and/or Tycho. Second (and related to the first issue), the \PItest test selection heuristics deciding which tests should be run first (see \secref{section:pitest}) could not find the tests due to dynamic loading of components.

\item \textbf{Cheshire.} [\url{http://github.com/AlFranzis/cheshire}] As the next option, we explored the possibility to automatically convert the project into a non-OSGI one during the build. Cheshire is a prototype tool that provides an interface for OSGI-compliant software to resolve and retrieve dependencies during compile time. After e-mail communication with the developer of Cheshire, it was clear that many extra recipes should be written for the various kinds of dependencies used within the industrial case. %
According to the developer's estimation, it would take weeks for someone not familiar with the details of the component. Even then, the lack of previous experience with a combination of \PItest and Cheshire made the final result unpredictable. Therefore, this solution was dismissed.

\item \textbf{MuUnit.}  [\url{https://code.google.com/p/muunit}] Finally, we tried to incorporate an OSGI-compliant tool to perform the mutation testing. The only prototype suitable for this task was MuUnit~\cite{Haschemi2010}. After a quick try-out it became clear that at the time of our analysis (which was September 2014) MuUnit was an early prototype able to run its analysis on simple projects only. Since then, there has been no development on this project, and as it stands, it can be considered an abandoned project. For this reason, this solution was dismissed.

\item \textbf{Others.} We considered two other options, Jumble [\url{http://jumble.sourceforge.net/}] and Javalanche [\url{http://www.st.cs.uni-saarland.de/mutation/}]. A cursory analysis of these tools and their documentation revealed that the OSGI components would cause similar problems as we had with \PItest, hence these options were dismissed as well.

\end{compactitem}

\RQsubdivision{Lessons Learned.} 
This feasibility study demonstrated that---contrary to common wisdom---it is not that easy to integrate mutation testing into a complicated build process. This is caused by the interference between the mutation testing (deeply coupled with the test infrastructure in order to speed up the process) and the product line configuration (with dynamic loading of test components). As often within software engineering, it is an accidental problem not an essential one~\cite{Brooks1987}. Indeed, if the development team of the tools under investigation would choose to do so, they could probably engineer a solution. Yet, at the time of analysis the OSGI headers were too deeply embedded in the case under investigation to be handled by the tools available. 

Due to aforementioned problems, Ali Parsai (the first author of the paper) developed and adapted a proof-of-concept tool called \LittleDarwin, described in \secref{section:tools}.  \LittleDarwin is explicitly designed to be loosely coupled to the test infrastructure, completely relying on the build system to run the tests. However, in doing so \LittleDarwin forsakes the speed-up enabled by deep analysis of the test infrastructure and fast mutation injection via byte-code manipulation. We successfully applied \LittleDarwin to perform the mutation testing on Segmentation. The problems we encountered using other tools were alleviated by the fact that \LittleDarwin itself does not run the tests, but rather, the build system does. Therefore, the build system fetches the OSGI dependencies, and creates and configures the test harness. By using \LittleDarwin to analyze Segmentation successfully, we demonstrated the feasibility of mutation testing within an industrial \CI environment.

\hypobox{\textbf{RQ1 Summary}\\ Due to the accidental complexity, byte-level mutation tools such as \PItest cannot easily be integrated into a complicated build process. Yet, if one decouples the mutation tool from the test infrastructure (thus relies on the build system to execute the tests) it is feasible to integrate mutation testing in a \CI setting. However, one does so at the expense of performance, i.e. mutation testing cannot be done as smartly and efficiently as the tightly coupled counterpart.}

\subsection*{\textit{RQ2. \RQtwo}} 

\RQsubdivision{Motivation.}
In a development environment where software changes frequently and in an incremental manner, having an accurate view on the quality of the test suite is a necessity. Mutation coverage is a strong contender for this role, whereas branch coverage is more widely used in industry.
On the one hand, mutation coverage has been demonstrated to subsume branch coverage~\cite{Offutt1996a,Li2009}. 
On the other hand, Gligoric et al. reports that among several coverage criteria, branch coverage is the best one to predict the mutation coverage of a test suite~\cite{Gligoric2013}. 
In this context, analyzing them together helps to determine whether or not mutation testing is capable of highlighting where branch coverage offers false confidence on the quality of the test suite. 

\RQsubdivision{Approach.}
Just as with RQ1, the unit of analysis is a class.  We collect branch coverage via \JaCoCo and mutation coverage via \LittleDarwin for one industrial system and four open source systems listed in \tabref{table:casestatistics}. For each class in the systems under study, we analyze the branch and mutation coverage, classifying them in the five categories shown in \tabref{table:categories}.  By focusing on code sections characterized by the difference between branch coverage and mutation coverage, we show how mutation coverage exposes additional weaknesses.
The similarity threshold we derived was $t=11$ for the industrial case and $t=7$ for the open source cases. We also calculate the Kendall correlation coefficient ($\tau_b$) and the p-value.

\begin{figure*}
	\centering
	\includegraphics[width=\linewidth]{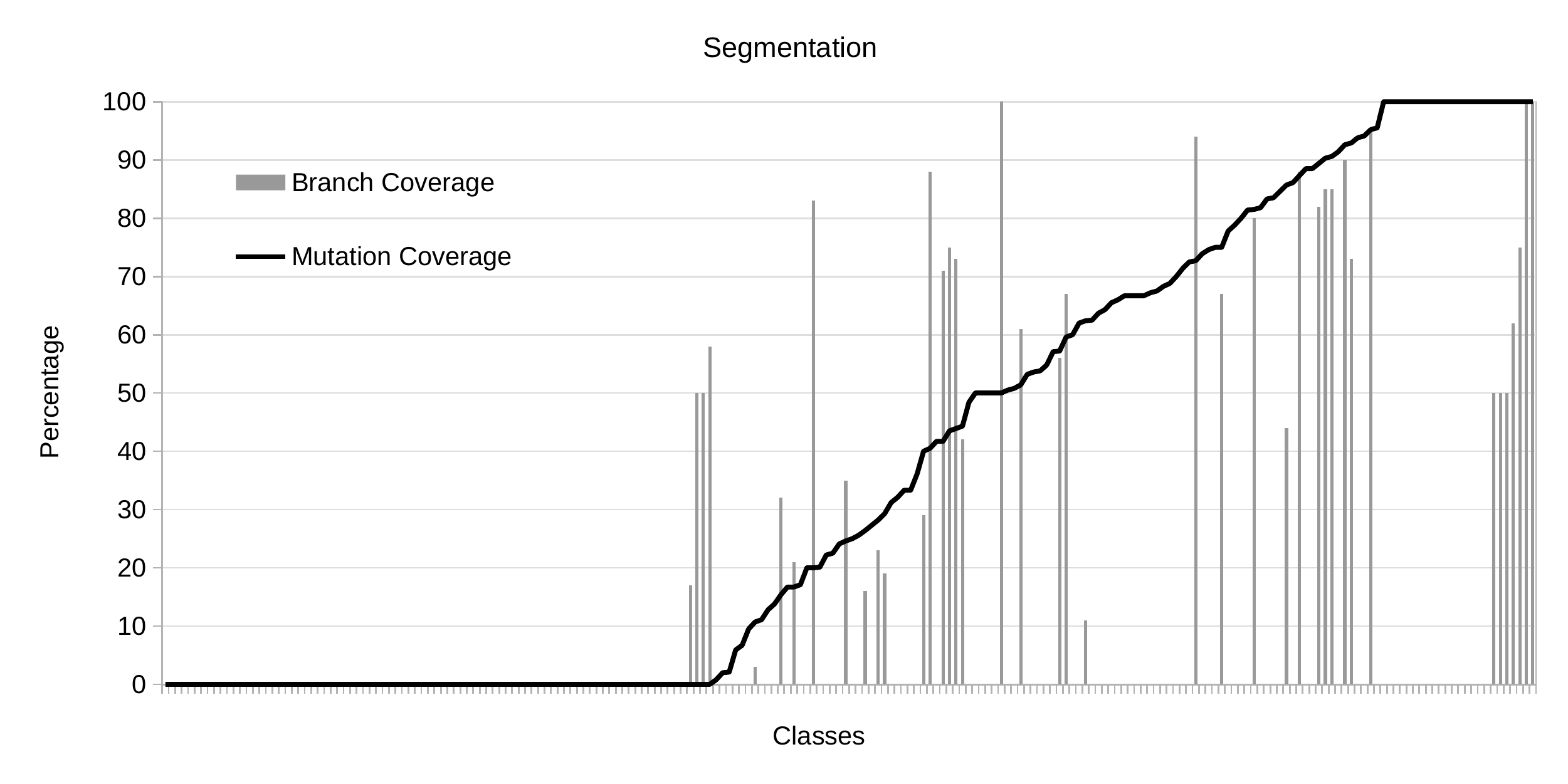}

{\footnotesize [The horizontal axis represents the classes sorted by mutation coverage; the vertical axis is the coverage percentage. Each bar represents the branch coverage for a class, and each point on the line represents the mutation coverage for a class.]}
	\caption{Visual comparison of branch coverage (\JaCoCo) versus mutation coverage (\LittleDarwin) on the industrial case}
	\label{fig:mutationcoveragechart}
\end{figure*}

\RQsubdivision{Findings for the Industrial Case.}
The industrial case has 212 classes. During the analysis, 4,955 of the 12,825 generated mutants were killed by the test suite, resulting in 38.6\% overall mutation coverage.  \figref{fig:mutationcoveragechart} shows the mutation coverage and branch coverage for each class. 
First of all, a large number of classes (48\%) have zero branch coverage and mutation coverage, illustrating the inadequacy of the test suite. This is true considering that we focus on the unit tests only; indeed, there was a suite of acceptance tests which did exercise most of the code but takes hours to execute. %
Secondly, there are several classes where the branch coverage and mutation coverage vary by a large margin.

The same data, organized according to the classification described in \secref{section:comparisoncriteria}, is reported in \tabref{table:experimentsummary}. 
The table lists the number of classes for each of the five categories, as well as Kendall correlation coefficient ($\tau_b$) and the p-value. First of all, we see that for 102 out of 212 classes (thus less than half) the branch and mutation coverage are the same for all practical purposes. For these classes, mutation testing does not provide additional value. Secondly, there are 8 classes in the category~\catLoBHiM, where the mutation coverage is larger than the branch coverage. Given the fact that \LittleDarwin includes ROR mutation operator, at first sight it is expected that wherever there is mutation coverage, it should guarantee branch coverage. However, with deeper analysis we found this not to be true because there are lots of multi-branched methods  that include  many arithmetic operations only in a few of branches. The tests often target only these branches due to their perceived importance. Consequently,  many of the large number of arithmetic mutants generated for these branches are killed, resulting in a high mutation coverage despite covering only few branches. Here, mutation testing provides extra confidence concerning the test suite; despite the low branch coverage, the test suite  has a high coverage over fault-prone areas of the code. For these classes, mutation testing provides additional value. Most interestingly, there are 8 classes in the category~\catHiBLoM, where the mutation coverage is smaller than the branch coverage. There the mutation testing reveals weaknesses in the test suite; the high branch coverage gives a sense of false confidence regarding the test suite, even though the mutation coverage shows that some covered branches are indeed not adequately tested.

The most extreme case in this respect is the class \textit{Discrete3DContour}, which has 88\% branch coverage yet only 41\% mutation coverage. For this class additional tests are needed, since the current tests cannot reveal injected faults. Most surprisingly (also apparent in \figref{fig:mutationcoveragechart}), there are 90 classes with zero branch coverage, yet some mutation coverage (Category \catNoB). %
One representative example is the class \textit{RegionGrowerNeighbours}; which has 0\% branch coverage yet all 22 generated mutants are killed by the \textit{VolumeGrowerTest} and \textit{RegionGrowerNeighboursTest}. Both tests indirectly verify the algorithms provided by \textit{RegionGrowerNeighbours}, and for this reason it is unlikely that the branch coverage is 0\%. Manual inspection confirmed that here as well it was the dynamic loading of components by means of OSGI which leads to the loss of execution traces and thus results in the miscalculation of the branch coverage in \JaCoCo (see \secref{section:jacoco}). Finally, there were four classes with zero mutation coverage yet significant branch coverage (58\%, 50\%, 50\% and 17\% respectively). %
There the branch coverage creates an even higher sense of false confidence: there is branch coverage, yet the tests fail to reveal any faults.

\begin{figure}
	\centering
	\includegraphics[width=0.9\linewidth]{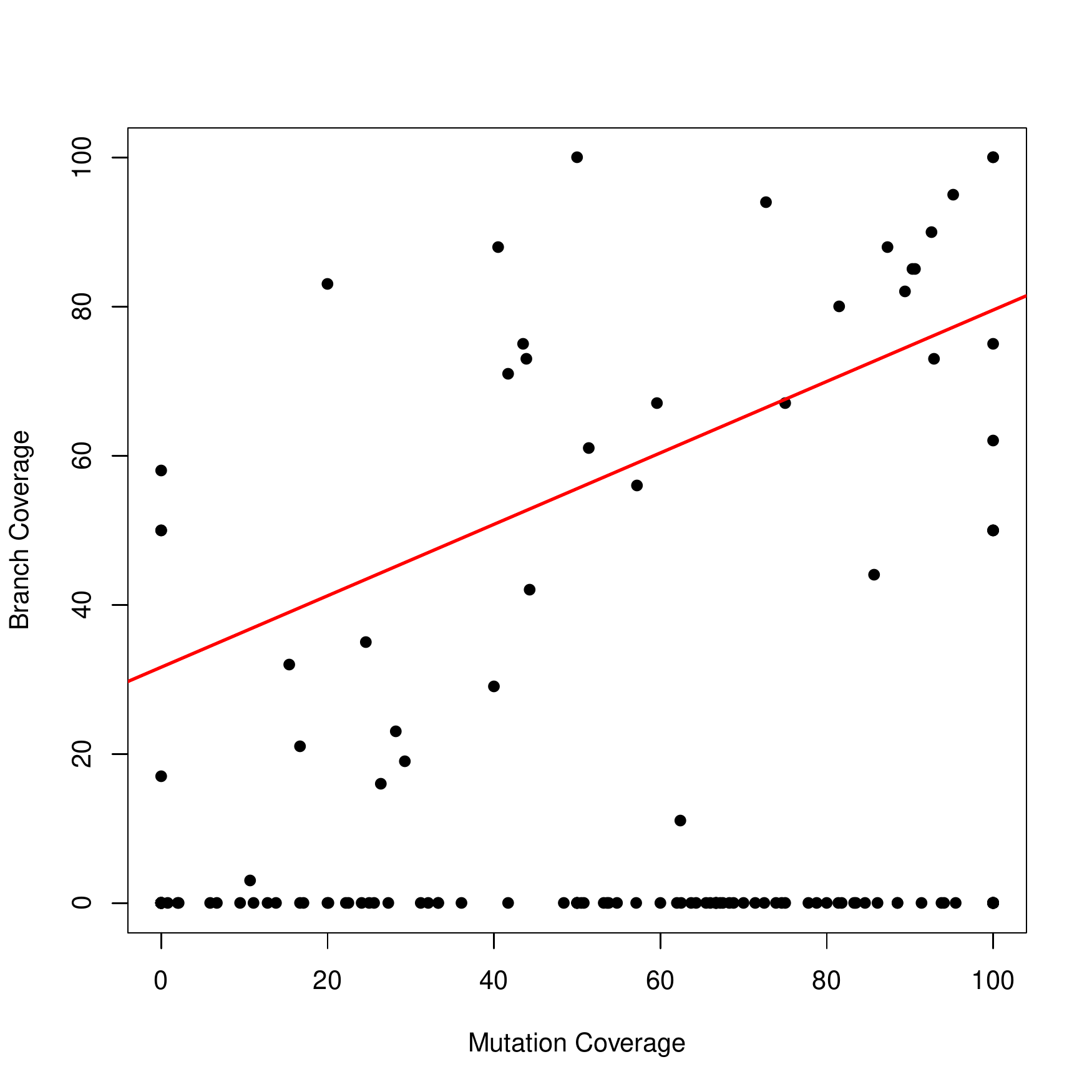}
	
	{\footnotesize [The horizontal axis represents mutation coverage; the vertical axis represents branch coverage. The red line is the linear regression line.]}
	\caption{Weak correlation between branch coverage (\JaCoCo) and mutation coverage (\LittleDarwin) on the industrial case}
	\label{fig:segmentation-scattered}
\end{figure}

\begin{table*}
	\centering
    \caption{Comparing branch coverage (\JaCoCo) versus mutation coverage (\LittleDarwin) on cases under study}	
    \label{table:experimentsummary}
    	\footnotesize
\begin{tabular}{|l|c|c|c|c|c|c|c|}
		\hline \multirow{2}{*}{\textbf{Case}} & \multicolumn{5}{c|}{\textbf{Categorization}} & \multicolumn{2}{c|}{\textbf{Correlation}} \\
		\hhline{~-------} & \catSimCov & \catLoBHiM & \catHiBLoM &\catNoB &\catNoM & \textbf{Kendall $\tau_b$} & \textbf{p-value} \\ 
		\hline \multicolumn{8}{|l|}{\textit{Industrial Case}} \\
		\hline Segmentation & 102 & 8 & 8 & 90 & 4 & 0.25 & $8.951\times10^{-6}$ \\ 
		\hline \multicolumn{8}{|l|}{\textit{Open Source Cases}} \\
		\hline Joda Time & 43 & 18 & 12 & 0 & 70 & 0.11 & $1.132\times10^{-1}$  \\ 
		\hline Apache Commons Codec & 27 & 3 & 7 & 0 & 0 & 0.31 & $1.755\times10^{-2}$  \\ 
		\hline jOpt Simple &  30 & 2  & 1 & 0  & 1 & 0.71 & $2.863\times10^{-6}$ \\ 
		\hline AddThis Codec & 9 & 9  & 8 & 0 & 0 & 0.53 & $2.146\times10^{-4}$ \\ 
		\hline\hline Total & 211 & 40 & 36 & 90 & 75 & \multicolumn{2}{c|}{ N/A } \\
		\hline
\end{tabular}

\end{table*}

\figref{fig:segmentation-scattered} shows the weak correlation between branch and mutation coverage values of all classes in the industrial case. While correlation cannot be ruled out (Kendall correlation of 0.25, p-value $<0.01$), it shows that branch and mutation coverage correlate weakly at best. %
This may be partially attributed to the 90 classes with zero branch coverage (the horizontal line of dots at the bottom of the chart), which is an anomaly caused by the dynamic loading of OSGI components. 

\begin{figure*}
	\begin{minipage}{\linewidth}
	\centering
	\includegraphics[width=\linewidth]{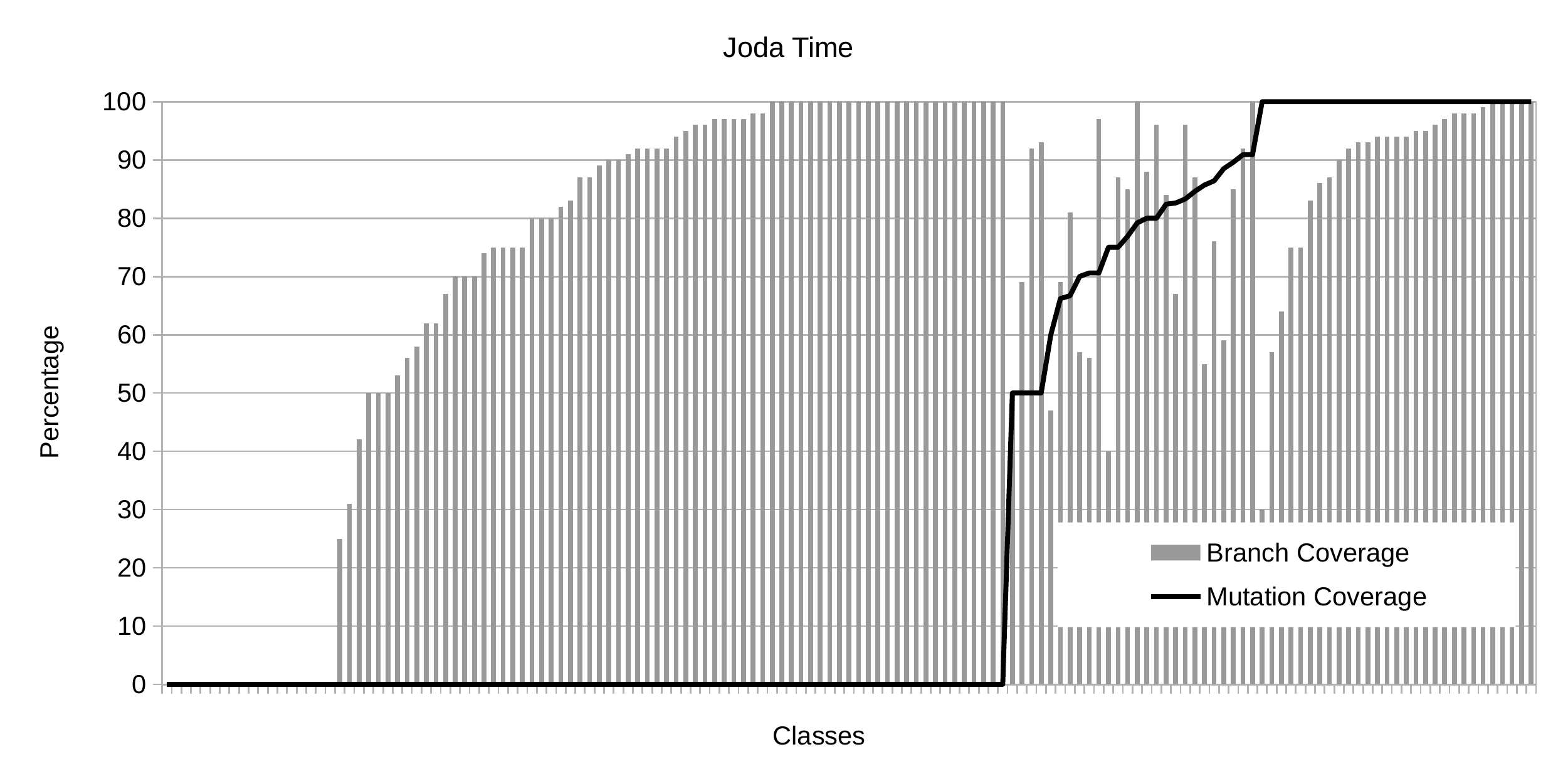}
	\end{minipage}
	\begin{minipage}{\linewidth}
			\centering
			\includegraphics[width=\linewidth]{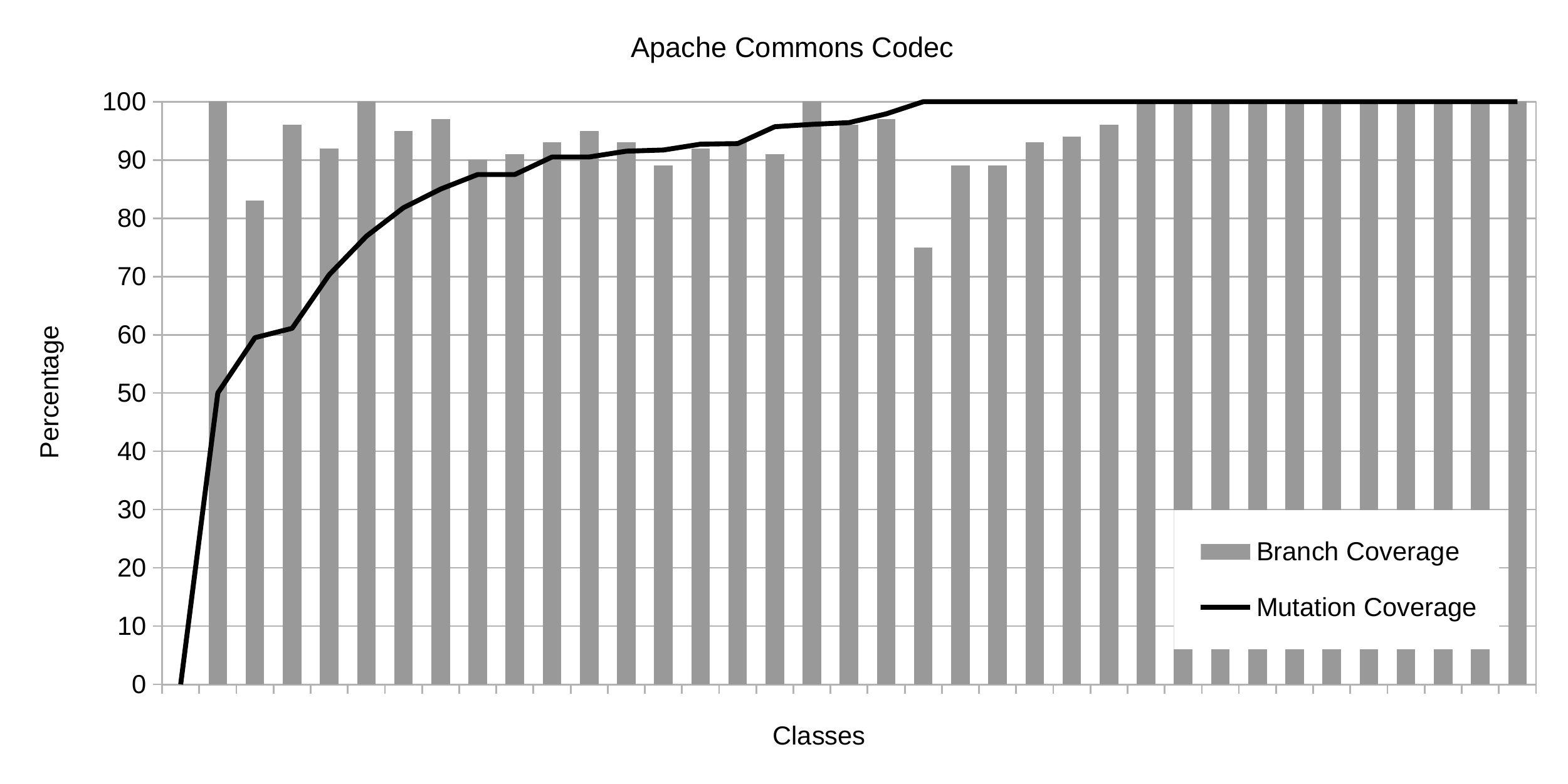}
		\end{minipage}
		
			{\footnotesize [The horizontal axis represents the classes sorted by mutation coverage; the vertical axis is the coverage percentage. Each bar represents the branch coverage for a class, and each point on the line represents the mutation coverage for a class.]}
				\caption{Branch coverage (\JaCoCo) and mutation coverage  (\LittleDarwin) at class level for the open source cases}
				\label{fig:barplotresults}
\end{figure*}
\begin{figure*}
\ContinuedFloat
	\begin{minipage}{\linewidth}
		\centering
		\includegraphics[width=\linewidth]{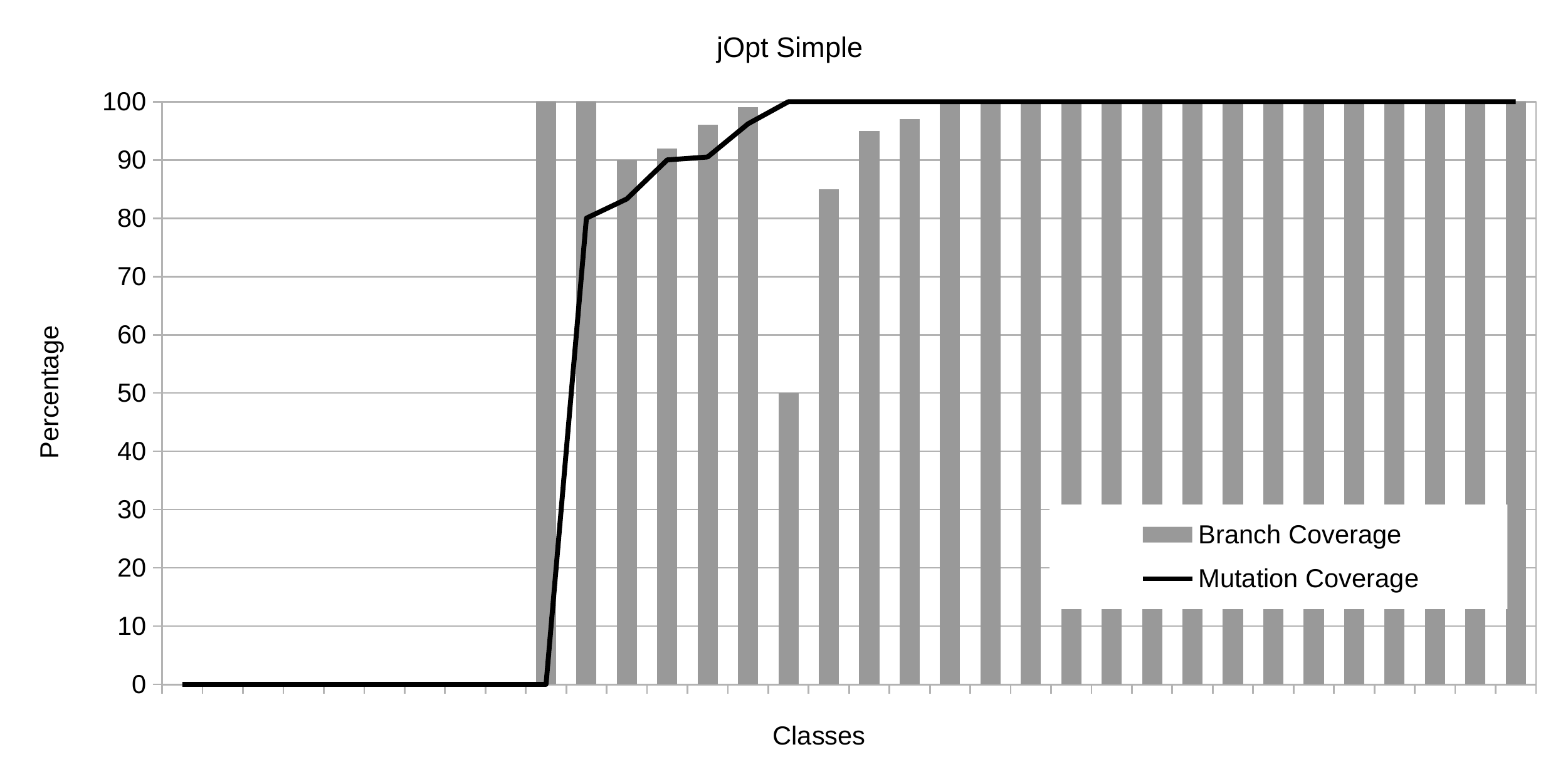}
	\end{minipage}
		\begin{minipage}{\linewidth}
			\centering
			\includegraphics[width=\linewidth]{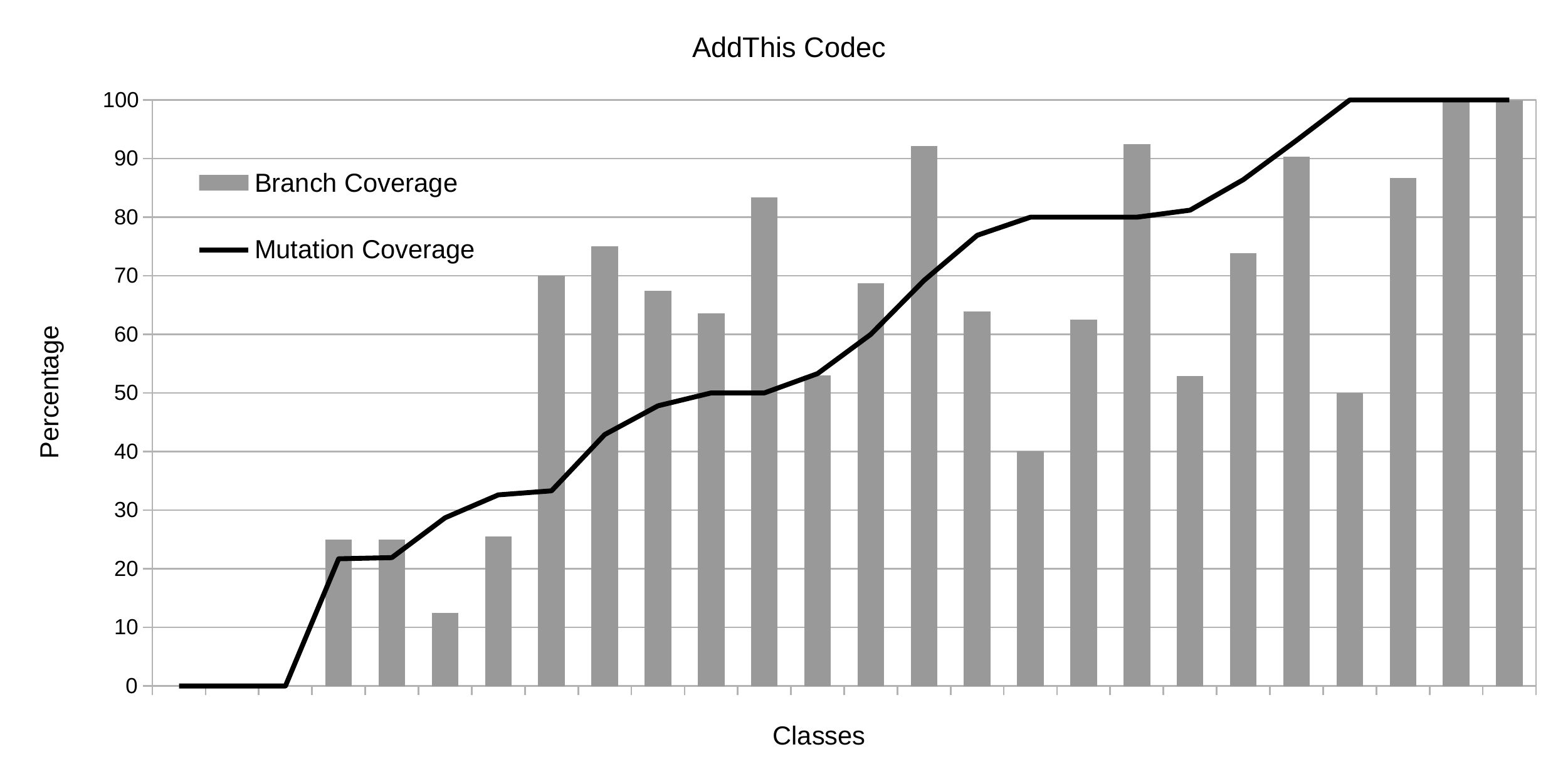}
		\end{minipage}

	{\footnotesize [The horizontal axis represents the classes sorted by mutation coverage; the vertical axis is the coverage percentage. Each bar represents the branch coverage for a class, and each point on the line represents the mutation coverage for a class.]}
		\caption{(Continued from previous page) Branch coverage (\JaCoCo) and mutation coverage  (\LittleDarwin) at class level for the open source cases}

\end{figure*}

\begin{figure*}
	\begin{minipage}{0.5\textwidth}
		\centering
		\includegraphics[width=\linewidth]{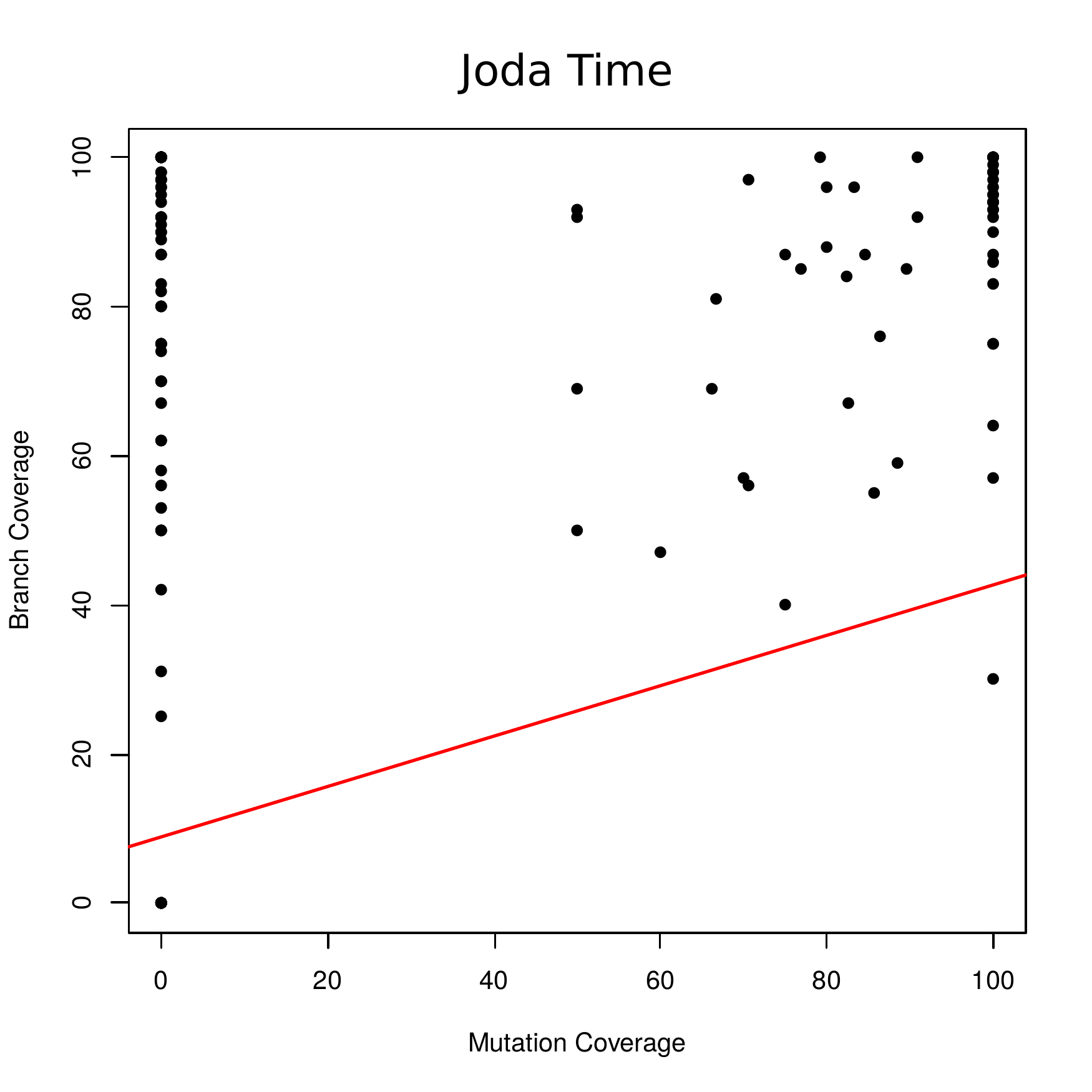}
	\end{minipage}
	\begin{minipage}{0.5\textwidth}
		\centering
		\includegraphics[width=\linewidth]{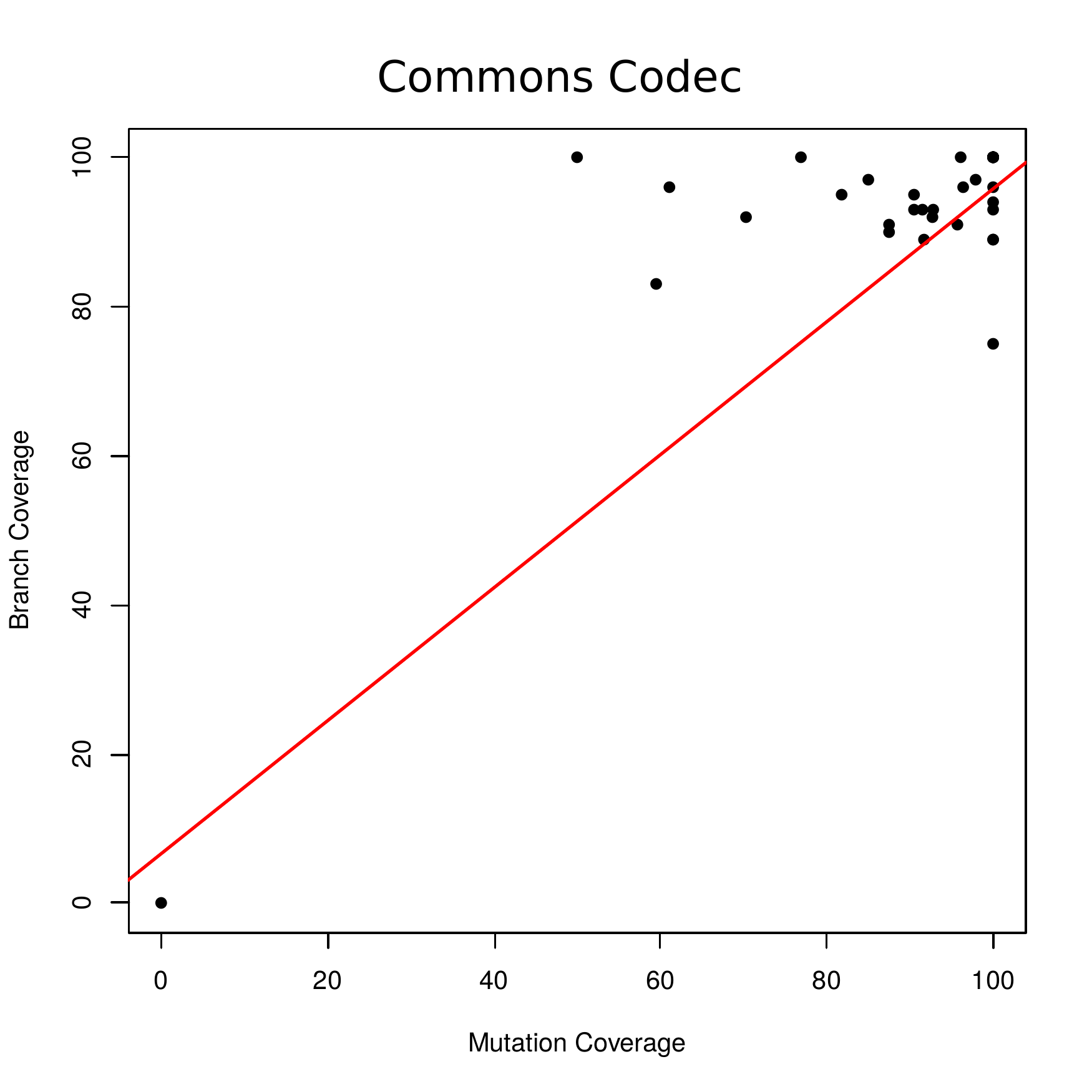}
	\end{minipage}
	\begin{minipage}{0.5\textwidth}
		\centering
		\includegraphics[width=\linewidth]{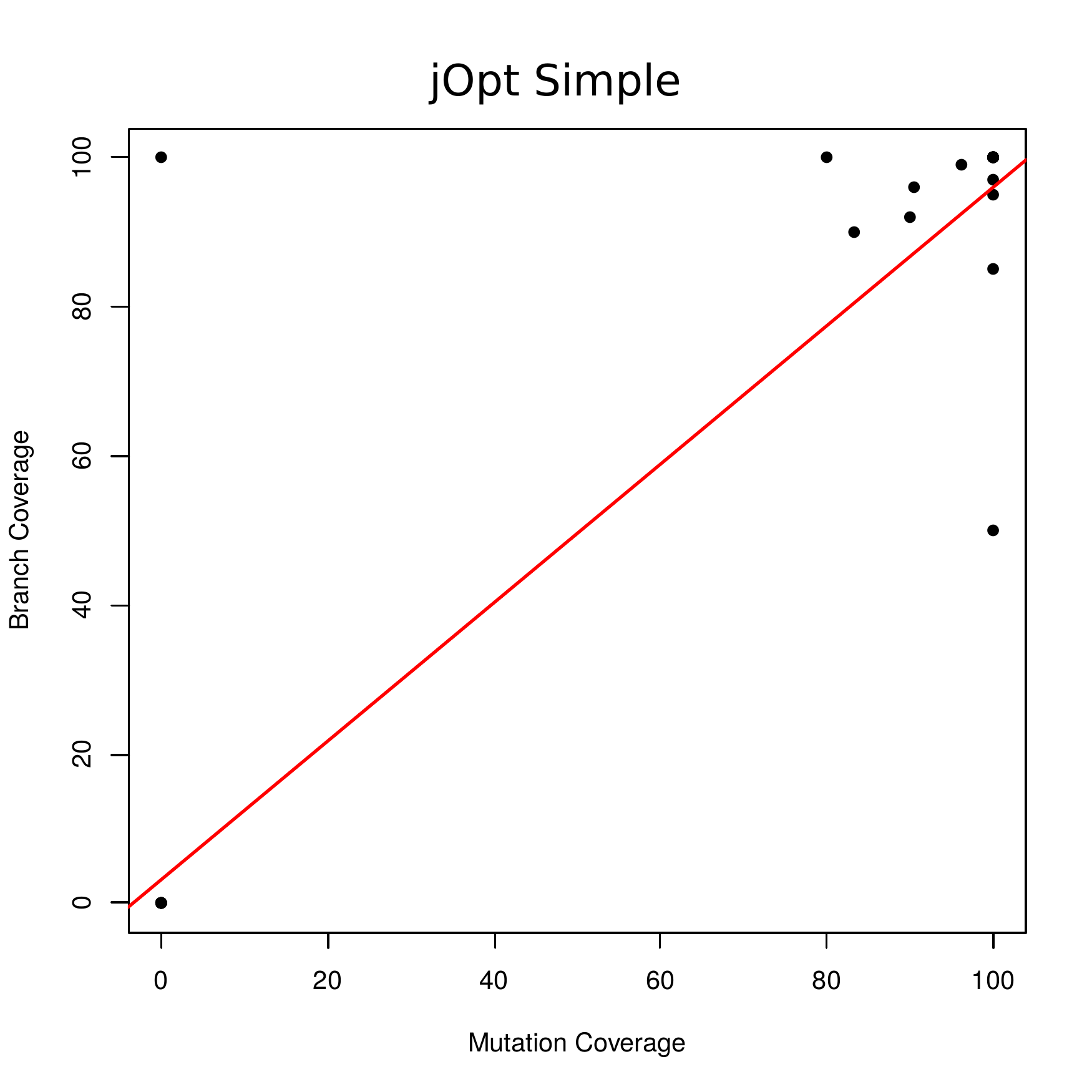}
	\end{minipage}
	\begin{minipage}{0.5\textwidth}
		\centering
		\includegraphics[width=\linewidth]{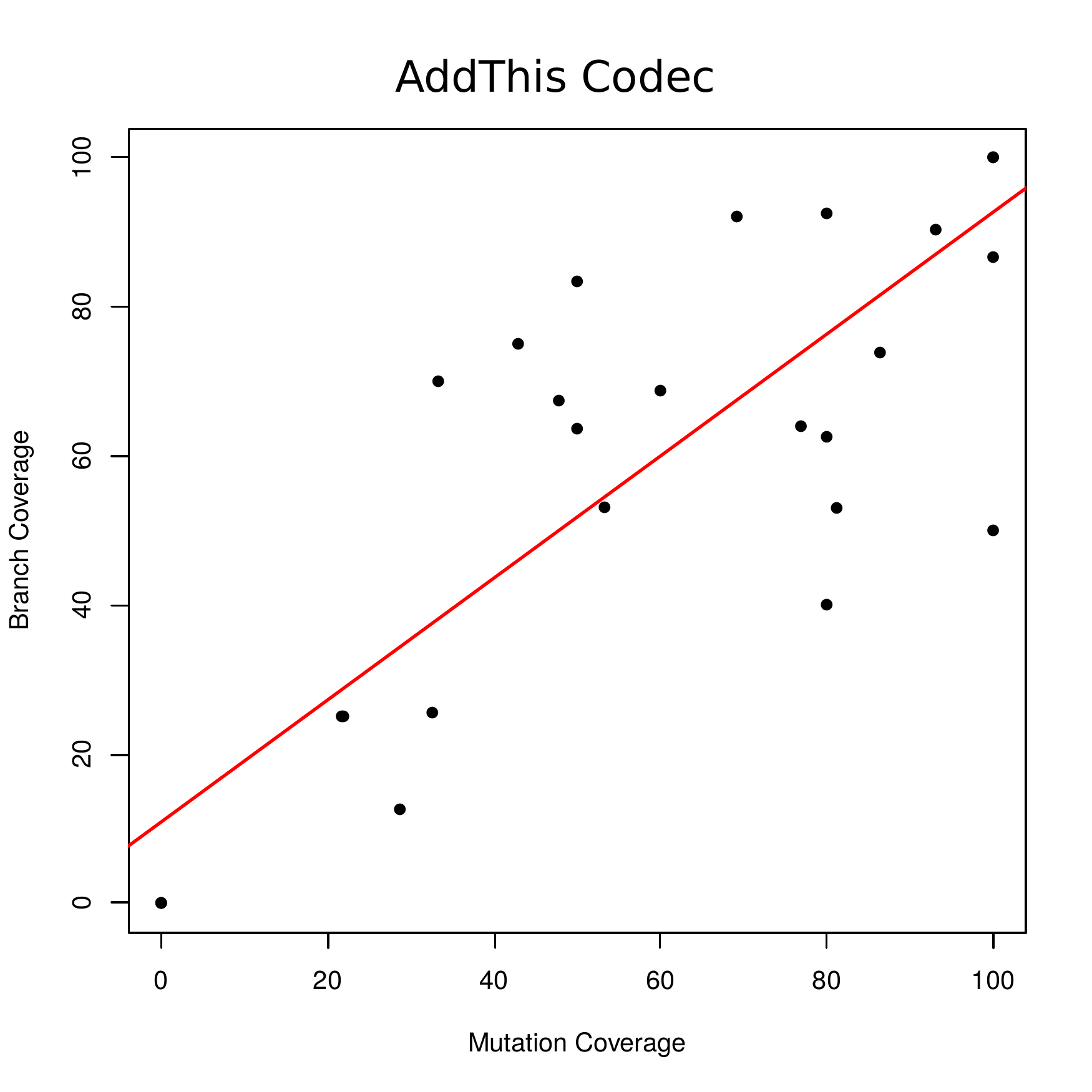}
	\end{minipage}
	
	\centering
	{\footnotesize [The horizontal axis represents mutation coverage; the vertical axis represents branch coverage. The red line is the linear regression line.]}
	\caption{Correlation between branch coverage (\JaCoCo) and mutation coverage (\LittleDarwin) on the open source cases}
	\label{fig:scatterresults}
\end{figure*}

\RQsubdivision{Findings for the Open Source Cases.}
\figref{fig:barplotresults} shows the mutation and branch coverage values for all classes in the open source cases. Most strikingly, the results of the analysis for open source cases are quite different than from the industrial case. This is quite apparent in case of \texttt{Joda Time}, where branch coverage is providing more information than mutation coverage. Moreover, the low number of classes with no coverage at all shows that the open source cases are more adequately tested compared to the industrial case. The same data, organized according to the classification described in \secref{section:comparisoncriteria}, is listed in \tabref{table:experimentsummary}. All four cases show some degree of branch coverage ($\catNoB = 0$); three of the four open source cases (\texttt{Apache Commons Codec}, \texttt{jOpt Simple} and \texttt{AddThis Codec}) also have some degree of mutation coverage  ($\catNoM <= 1$). Here as well, \texttt{Joda Time} is the outlier ($\catNoM = 70$). Manual inspection revealed that a lot of the classes in \texttt{Joda Time} were implemented without any statements that could be mutated by the mutation operators of \LittleDarwin (see \secref{section:littledarwin}). For example, some data classes include only variables and getter/setter methods, and therefore no mutant is generated for them. 
This illustrates that it might be worthwhile to expand the mutation operators beyond the reduced set listed in \tabref{table:reducedsetofoperators}. 

Looking at the column \catSimCov, we see that for most of the classes (ranging from 30\% to 88\%) the branch and mutation coverage are the same for all practical purposes. Thus for more than half of the classes, mutation testing does not provide additional value. This is quite different from the industrial case, where less than half of the classes had similar coverage values. Nevertheless, there are a significant number of classes in the category~\catLoBHiM, where the mutation coverage is larger than the branch coverage. Furthermore, there are also a significant number of classes in the category~\catHiBLoM, where the mutation coverage is smaller than the branch coverage. For three of the four cases, the values in column \catHiBLoM are larger than the values in column \catLoBHiM and here as well \texttt{Joda Time} is the exception. Thus, although much less than in the industrial case, there are still a significant number of classes where mutation tests reveals additional weaknesses.

Analyzing the correlation between branch and mutation (i.e. columns ``Kendall $\tau_b$'' and ``p-value'')  we see that the correlation between branch and mutation coverage is rather poor. \figref{fig:scatterresults} 
provides insight into the lack of correlation. For each of the cases the dots in the scatter plot are in distinct regions, hence the coverage values depend a lot on the particular context of the case under investigation. 

\RQsubdivision{Lessons Learned.} Comparison of branch coverage and mutation coverage seen in  \tabref{table:experimentsummary} shows that these measures agree only for 47\% of analyzed classes. In 9\% of classes, we observed that the density of the mutants were much higher in few branches in the code, and thus despite a low branch coverage, the mutation coverage is much higher. Conversely, in 8\% of cases where  branch coverage is high and mutation coverage is low, the quality of the test oracles are in question. In the remaining 36\% of classes, the peculiarities of the code cause problems in calculation of either metric: In case of branch coverage, complicated structure of Segmentation  means that it cannot be accurately calculated, and this is in line with the observations of Tengeri et al.~\cite{Tengeri2016}. In case of mutation coverage, the abundance of small data classes or interfaces and stubs in Joda Time means that a more extensive set of mutation operators is required. All things considered, such analysis on the  weaknesses in the   tests cannot be done without mutation testing, and it is clear that using branch coverage alone can mislead a developer about quality of the tests.

\hypobox{\textbf{RQ2 Summary}\\
For all the cases under investigation, we discovered that significant number of classes where mutation testing reveals additional weaknesses compared to branch coverage. However, the added value of the mutation testing is context-dependent and varies between the cases we investigated. This warrants further research into the nature of lower coverage values both for branch coverage and mutation coverage.}

\subsection*{\textit{RQ3. \RQthree}}

\RQsubdivision{Motivation.} The results of RQ2 shows that mutation testing provides valuable complementary information over branch coverage. 
However, in order to avoid issues connected with specifics of the system architecture, the results of RQ1 suggest that we need to do so with mutation testing tools loosely coupled to the test infrastructure, thus inherently slower. This needs to be mitigated if mutation testing is to be incorporated into development environments with small, frequent, and incremental changes to the software.
For this reason, in this RQ we want to tackle the ``time concern''; 
demonstrating that (i) the performance overhead induced by mutation testing can 
be reduced to meet industrial time constraints and (ii) sampled mutation testing preserves the information presented in RQ2.
As a realistic scenario we consider a team that starts working from Monday at 8am till Friday at 6pm. Here,
we want to verify whether a complete mutation testing could run once a week during the week-end and right before the sprint meeting scheduled on Monday morning. In this context, we define an acceptable level of performance overhead as
a mutation testing job that runs in up to 62 hours, namely between Friday 6pm to Monday 8am. If the full mutation testing cycle takes longer, we use \emph{mutant sampling} to reduce the number of mutants injected into the system, because it has been demonstrated that even with a sample size as low as 50\% mutation tests still provides reliable results~\cite{Zhang2010}. 

\begin{figure}
\centering
\includegraphics[width= \linewidth]{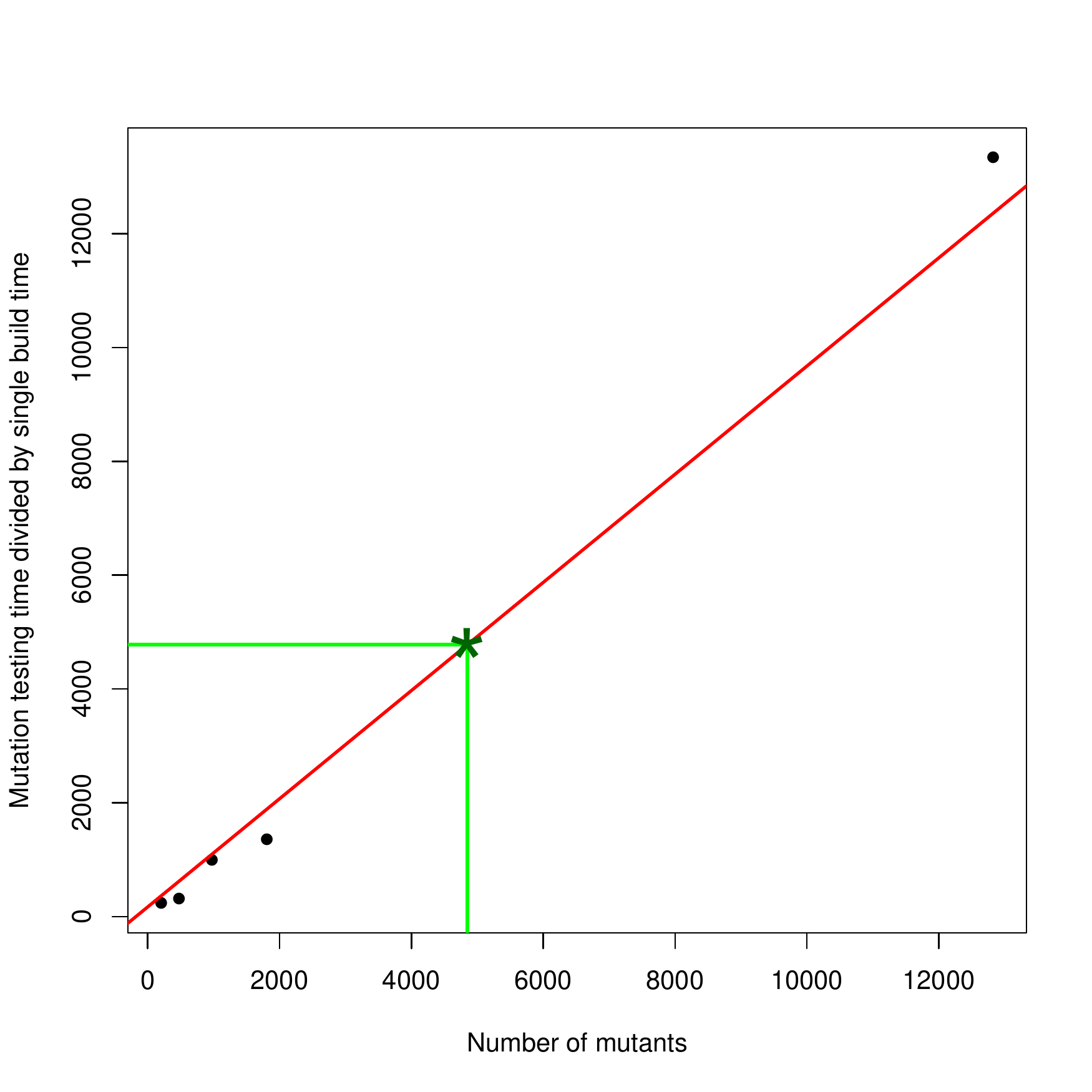}
\caption{Linear relationship between number of mutants and total analysis time}
\label{fig:timing-scatter}
\end{figure}

\begin{figure}
    \centering
    \includegraphics[width=\linewidth]{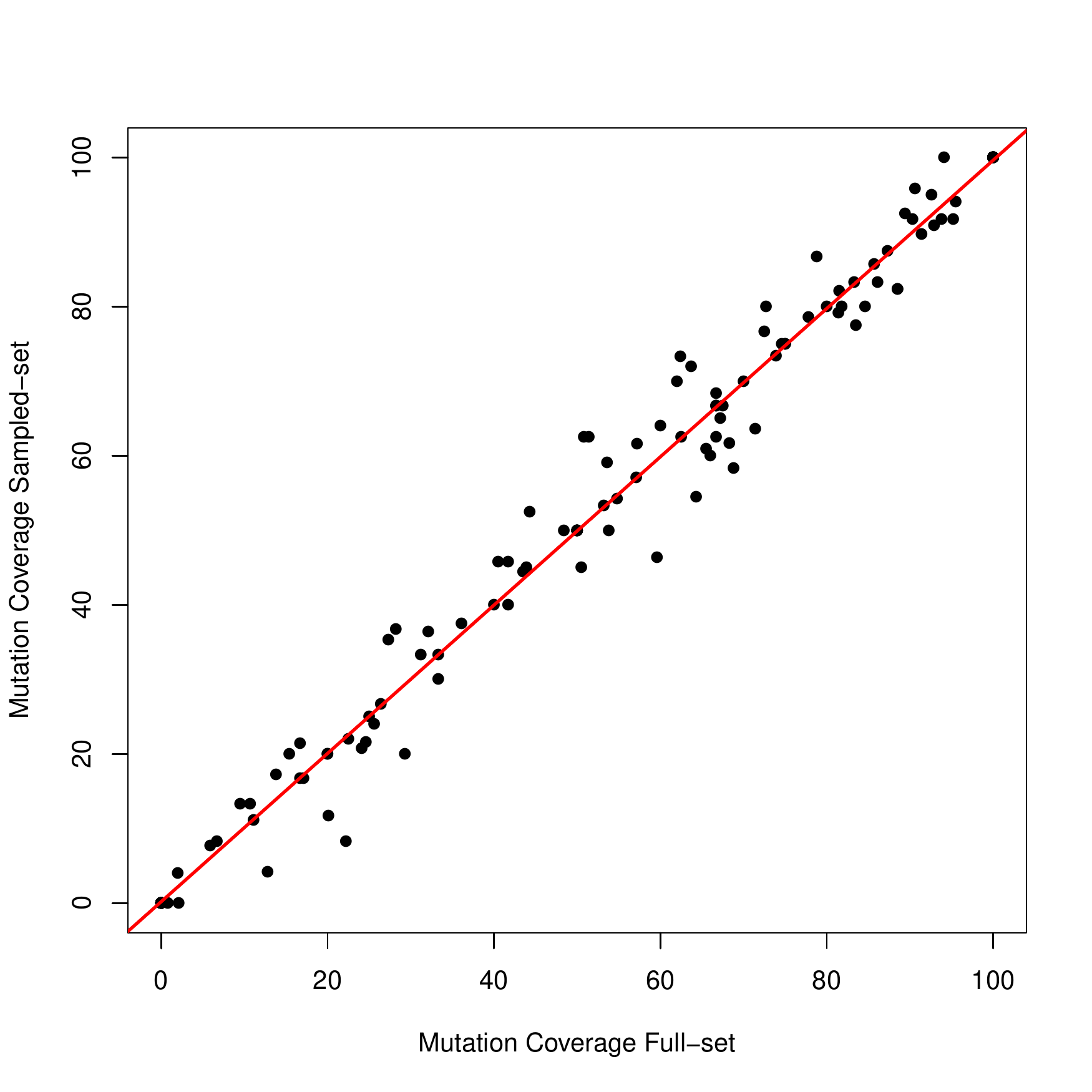}
    \caption{Correlation of full mutation coverage against mutation coverage with weighted sampling}
    \label{fig:mutationcoveragesampled}
\end{figure}

\begin{figure*}
    \centering
    \includegraphics[width=\linewidth]{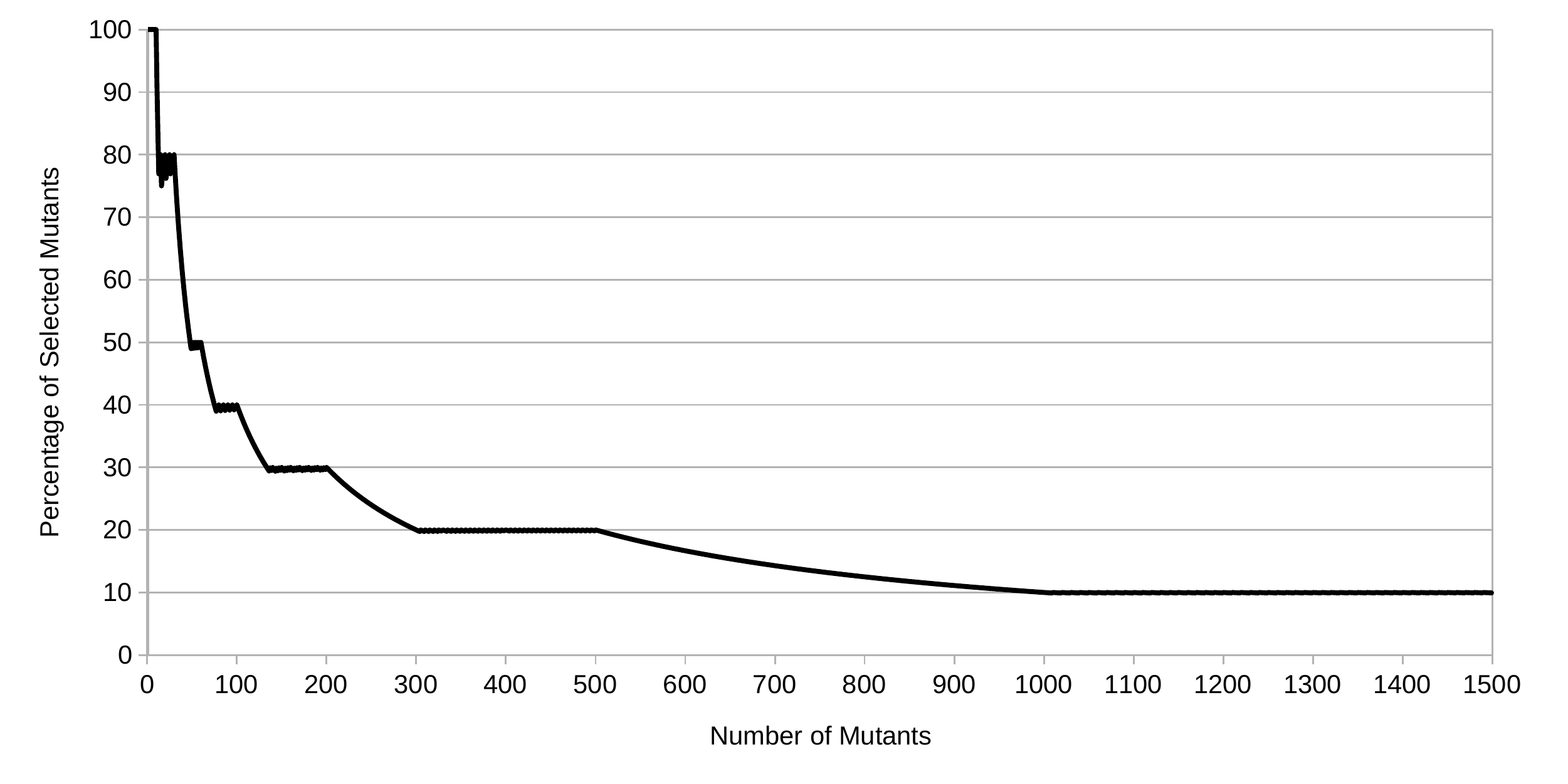}
    
    {\footnotesize [The horizontal axis shows the number of mutants generated for a class;\\ the vertical axis shows the percentage of selected mutants.]}
    \caption{The distribution of the \emph{weighted mutation sampling}}
    \label{fig:mutantsamplingselection}
\end{figure*}

\RQsubdivision{Approach.}
We first measure the performance overhead induced by a full mutation testing of the industrial case (injecting 12,825 mutants for 38K lines of code). Here we calculate the average time for a single iteration by dividing the total time for the process by the number of mutants. Next, we estimate the sample size based on the average single iteration time so that the total  analysis time would drop below the 62 hour maximum value. We then compare the coverage values for the full mutation testing and the sampled mutation testing with the same categories as RQ2.

To settle the sampling rate, we used the following procedure. We first analyze the relationship between the number of mutants, and the time required for the analysis. In principle, this should be a linear relationship, which is confirmed in \figref{fig:timing-scatter}. Via a linear regression ($\rho = 0.9992$,  p-value $=0.000026$) we achieve the slope of the regression line as 0.9506 with an offset value of 169.5 (Equation~\ref{eq:timemutants}). Using this equation, we can estimate the upper limit for the number of mutants. The total time needed to perform the analysis is $= 62$ hours; a single iteration takes $46$ seconds, thus we get as an upper limit for the number of mutants $4,780$. %

\begin{equation}
\label{eq:timemutants}
\resizebox{0.9\linewidth}{!}{$Number\ of\ mutants = 169.5 + 0.9506\times\frac{Total\ analysis\ time}{Time\ for\ a\ single\ iteration}$}
\end{equation}

\begin{table*}
    \caption{Comparing branch coverage (\JaCoCo) versus mutation coverage (\LittleDarwin)---full coverage + weighted sampling)}	
    \label{table:experimentsampling}	
    \footnotesize
    \centering
    \begin{tabular}{|c|c|c|c|c|c|c|c|}
        \hline \multirow{2}{*}{\textbf{Case}} & \multicolumn{5}{c|}{\textbf{Categorization}} & \multicolumn{2}{c|}{\textbf{Correlation}} \\
        \hhline{~-------} & \catSimCov & \catLoBHiM & \catHiBLoM &\catNoB &\catNoM & \textbf{Kendall $\tau_b$} & \textbf{p-value} \\
        \hline Industrial Case (full) & 102 & 8 & 8 & 90 & 4 & 0.25 & $8.951\times10^{-6}$ \\ 
        \hline Industrial Case (sampled) &  102 & 10 & 9 & 87 & 4 & 0.26 & $4.106\times10^{-6}$  \\ 
        \hline 
    \end{tabular}
    
\end{table*}

To guarantee that the classes with smaller set of mutants are still represented in the sample set, we used a procedure called \emph{weighted mutation sampling}~\cite{Parsai2016}, and we performed the random sampling at class-level rather than project level. 
The weights are chosen based on the size of the mutant set for each class, thus reduces the masking effect of classes with larger set of mutants in the final coverage score. This way, our sampling method randomly selected 4,448 mutants out of  12,825 generated mutants (34.7\%). The percentage of mutants selected based on the size of the mutant set is shown in \figref{fig:mutantsamplingselection}. 

\RQsubdivision{Findings.}
The full mutation testing of the industrial case (i.e. the Segmentation component) takes 163 hours to complete; way more than the established upper limit of 62 hours. 
Via the weighted sampling method, we reduced the number of mutants from $12,825$ to $4,448$ ($34,6\%$). With this reduction, the time required to perform the mutation testing dropped to 58 hours (almost one-third of the full analysis) well under the established upper limit. 

Out of the $4,448$ mutants in the sampled mutant set, $1,721$ were killed by the test suite, resulting in a 38.7\% overall mutation coverage. The difference between the overall coverage calculated from the sampled set and the one calculated from the full set is only 0.1\%. \tabref{table:experimentsampling} classifies the differences in mutation scores using the categories in  \tabref{table:categories}. The first row shows the values for the weighted mutation sampling, the second row (copied from \ref{table:experimentsummary}) show the values for full mutation testing. We see that the number of classes in each category remains roughly the same, thus confirming that sampling does not diminish the confidence in the validity of mutation coverage. This is confirmed in \figref{fig:mutationcoveragesampled}, which shows the correlation between the full mutation coverage values and the sampled ones; they do indeed have a very strong correlation ($\rho = 0.99$,  p-value $<0.00001$).

\hypobox{\textbf{RQ3 Summary}\\ When a full mutation testing exceeds the time limitations imposed by a \CI setting (i.e. an analysis once a week during the week-end), 
we can use weighted mutation sampling to reduce the performance overhead and at the same time preserve an accurate estimation of the mutation coverage.}

%% file: 06-ThreatsToValidity.tex
\section{Threats to Validity}
\label{section:threats}

We now identify factors that may jeopardize the validity of our results and the actions we took to reduce or alleviate the risk. 
Consistent with the guidelines for case studies research (see~\cite{Runeson2008}) we organize them into four categories.

\subsubsection*{Internal Validity.} 
Threats to internal validity focus on confounding factors that can influence the obtained results. In this study, this mainly concerns equivalent mutants and a limited set of mutation operators. Because of the large number of generated mutants, it is very difficult to check for equivalent mutants in the final generated results due to the amount of manual labor needed to find and remove such mutants. Nevertheless, since equivalent mutants would add to false positives, they would be discovered when the developers are trying to create new tests or improve the available tests by referring to the information acquired from mutation testing. As is common practice in today's research, we just accept the risk~\cite{Fawcett2006}.

Another threat stems from the limited set of mutation operators used in \LittleDarwin. The addition of more mutation operators does not impact the results of our industrial case study, where mutation coverage (even with the limited set of mutation operators) produced more information than branch coverage. However, addition or removal of mutation operators can have an adverse effect on the quality of mutant sampling, since there are mutation operators that can produce a large number of redundant mutants (e.g. Null-Type mutation operators~\cite{Parsai2019}), and thus affect the distribution of the sampled mutants. To identify such effects in practice, further case studies on industrial software is required; nonetheless, we did not pursue this line of research in this study.  

A large majority of the produced mutants are indeed redundant, and this affects the stability of mutation coverage as a metric~\cite{Papadakis2016}. In order to remove the redundant mutants, mutant subsumption relationships need to be used~\cite{Kurtz2014}. However, determining these relationships is a very difficult task at large scale. There are no available tools to our knowledge that performs static subsumption analysis on Java programs of this scale. In addition, dynamic subsumption analysis requires a very high quality test suite to be accurate. Therefore in case of non-adequate test suites of our subject projects, the dynamic subsumption relationships are unreliable to detect redundancy among mutants~\cite{Ammann2014}. For this reason, we did not filter redundant mutants in this study.

During the course of the study we discovered that \JaCoCo does not report branch coverage correctly in some cases. This phenomenon has been already documented in Tengeri et al. \cite{Tengeri2016,Blondeau2017}. Despite this fact, we decided to use the results as is for two reasons: first, we were informed by the developers of the industrial project that no other tool was capable of integration with their environment, and second, the results of the tool were used as is for the decision making process regarding the testing of the software. Given the fact that \JaCoCo is one of the most commonly used tools in maven builds, we believe its weaknesses are a reflection of the difficulties in computing branch coverage in practice, and therefore worth studying.

\subsubsection*{Construct Validity.} Threats to construct validity focus on how accurately the observations describe the phenomena of interest. This research is driven by RQ2 where we compare test coverage provided by two different tools. To minimize the risk on making wrong observations, we compare the test coverage in two different ways, once using the categories in \tabref{table:categories} and once via the Kendall's $\tau_b$ coefficient as described in statistic handbooks~\cite{Abdi2007,Agresti2010}. Moreover, we manually inspect certain results, especially outliers. 

In order to evaluate mutant sampling process, a common procedure is to create many smaller subsets of a test suite, and compare the mutation coverage obtained from sampled mutants and the mutation coverage obtained from all mutants for all of these smaller subsets of the test suite. Despite our attempts, this proved not to be possible in the industrial case. Because of the interdependencies between the tests, complicated setup process of the testing harness, and the use of shared resources, the order in which the tests are included and executed are important for a successful execution of the test suite. Therefore, it is not possible to randomly generate subsets of the original test suite. For this reason, we omitted this kind of analysis in our study.

\subsubsection*{Reliability.} Threats to reliability validity correspond to the degree to which the result depends on the tools used. The most important threat to validity concerns the tool \LittleDarwin implemented by first author. %
Compile errors and errors in tests can affect the final results, since \LittleDarwin checks only if the build process has failed or not, and it does not go further to determine the reason for the failure. This was addressed by inspecting the output of the build system. In this process, 107 invalid %
mutants were detected. These mutants could not be compiled, and were excluded from the final results.  Another threat to reliability validity is the fact that the data gathered by \JaCoCo might not be accurate (especially due to dynamic loading of components), and therefore the conclusions based on the comparison between branch coverage and mutation coverage might not be accurate enough. However, since \JaCoCo is the tool that is being used in the structure of Segmentation component to acquire this information in the first place, the conclusions are still relevant by challenging the previously held beliefs about the system. Because of active development of \JaCoCo and its popularity as an integral part of \CI systems, it is debatable  whether better accuracy can be achieved.   

\subsubsection*{External Validity.} Threats to external validity correspond to the generalizability of our results. Since this study was performed only on a software running on a single platform with a specific target language and a specific \CI environment. Therefore, the results are certainly not representative for all possible industrial systems. However, it provides an outlook on the feasibility of applying mutation testing in an industrial environment, especially concerning the challenges we faced and workarounds we performed. Nevertheless, we partially addressed the generalizability by extending the analysis regrading RQ2 on other open source cases. By comparing these results with those of the industrial case, we determined the situations  in which our conclusions can be generalized.

%% file: 07-RelatedWorks.tex
\section{Related Work}
\label{section:relatedwork}

There are several studies that assess the effectiveness of test coverage metrics. Offutt and Voas~\cite{Offutt1996a} use generated test cases for Fortran, and conclude mutation coverage subsumes condition coverage techniques. Li and Offutt~\cite{Li2009} compare mutation coverage with edge-pair, all-uses, and prime path coverage. They use hand-seeded faults and manually developed test cases in their comparison, and conclude that mutation coverage is more effective in detecting faults, and requires a smaller test suite to be satisfied. Gligoric et al.~\cite{Gligoric2013} compare non-adequate test suites by using several coverage criteria, and conclude that branch coverage is the best predictor of mutation coverage. Gopinath et al.~\cite{Gopinath2014} compare coverage criteria that is available to developers on a large set of open source cases, and conclude that statement coverage, and not branch coverage, is the best predictor of mutation coverage.   

Literature is clear on the dangers of using code coverage as a threshold for quality of the test suite. Marick~\cite{Marick1999} points out a scenario that relying solely on the code coverage metrics could result in faults not being detected. Inozemtseva and Holmes~\cite{Inozemtseva2014} compared decision coverage, modified condition coverage, and statement coverage on large subjects using generated test cases and mutation coverage as the effectiveness criteria, and concluded that while code coverage is good for identifying under-tested parts of the subject, it should not be  used as a quality target. Aaltonen et al.~\cite{Aaltonen2010} reach the same conclusion in their analysis comparing mutation coverage and code coverage metrics in assessment of students' skills. They propose adoption of both metrics in order to have an accurate assessment of the test suite quality. Smith and Williams~\cite{Smith2009,Smith2009a} studied the effects of using mutation testing to augment a test suite. They concluded that developing new test cases that increase the mutation coverage also increases branch coverage, and statement coverage of the test suite. They also conclude that the inclusion of new mutation operators is less important than the speed and efficiency of mutation testing process. 
Andrews et al.~\cite{Andrews2006} validate the use of mutation testing as a benchmark for other coverage criteria using an industrial case with known faults, and conclude that not only mutation testing can be used in a research context, but also in a practical context, it can be used as a threshold to develop new test cases. Li et al.~\cite{Li2015} use industrial cases written in Ruby and demonstrate that using mutation testing still adds value to a test suite that has 97\% statement coverage.

%% file: 69-Conclusions.tex
\section{Conclusion and Future Work}
\label{section:Conclusions} 
With the increasing interest in \CI and development environments where changes to the code occur in small and frequent steps, developers rely  fully automated tests to find faults as early as possible. Therefore, they require to continuously monitor the coverage of their automated test suites.
Unfortunately, the state of the practice is reluctant to adopt strong coverage metrics (namely mutation coverage), instead relying on weaker kinds of coverage (namely branch coverage). We argue that there are three issues for this reluctant attitude towards mutation testing: (a) the complexity of continuous build environments and (b) the perception that branch coverage is ``good enough''; (c) the performance overhead during the build. 
Consequently, we set out to investigate the pros and cons that arise when adopting mutation testing in an industrial continuous integration setting, namely the Segmentation component of the Impax ES medical imaging software used by Agfa HealthCare. The Impax ES system is configured as a product-line released in two main variants (production or prototype), with a few minor variants for the target hardware platform. The extensive use of (dynamic) OSGI headers implies a complicated build process where the \Maven plug-in Tycho [\url{https://eclipse.org/tycho/}] is used to fetch dependencies, compile source files, and run the test suite. This lead us to pursue the following research questions.

\noindent
\textbf{RQ1:} \emph{\RQone}
\begin{compactitem}
\item Byte-level mutation tools such as \PItest cannot be easily integrated into a complicated build process. Yet, if one decouples the mutation tool from the test infrastructure (thus relies on the build system to manipulate the tests) it is feasible to integrate mutation testing in a \CI setting. However, one does so at the expense of performance; that is to say mutation testing cannot be implemented as smartly and efficiently as the tightly coupled counterpart.
\end{compactitem}

\noindent
\textbf{RQ2:} \emph{\RQtwo}
\begin{compactitem}
\item We discovered that mutation coverage reveals additional weaknesses compared to branch coverage in all of the investigated cases. More specifically, there were several classes which had a high branch coverage yet a low mutation coverage, hence in such situations branch coverage gives a false sense of confidence.
However, the added value of the mutation coverage is context dependent and varies between the cases we investigated. This warrants further research into the nature of weak coverage values both for branch  and mutation coverage. Especially because  here too we noticed that dynamic loading of components interferes with calculating branch coverage.
\end{compactitem}

\noindent
\textbf{RQ3:} \emph{\RQthree} 
\begin{compactitem}
\item When a full mutation testing exceeds the time limitations imposed by a \CI setting (i.e. an analysis once a week during the week-end), we can reduce the performance overhead by means of weighted mutation sampling without sacrificing the fault detection capability.
\end{compactitem}

\noindent
The contributions of this research are fourfold. First, we report the challenges that arise when mutation testing is integrated in a continuous integration tool of a real industrial case. Second, we adapt and use our mutation testing tool called \LittleDarwin to overcome these challenges. 
Third, we perform a joint analysis of mutation and branch coverage on cases with non-adequate test suites. 
For this analysis we used four open source cases (totaling close to 40 thousand lines of code with varying degrees of test coverage) and one industrial case (with more than 38 thousand lines of code with limited unit test coverage).  
We demonstrate that in cases without full branch test adequacy, mutation testing does not subsume (as expected) branch coverage. Yet, it provides complementary information  that can be exploited for both  determining the fault detection ability of the test suite, and forming a long-term plan to improve it. 
We also did not find sufficient evidence to support previous conclusions in literature regarding the relation of branch coverage and mutation coverage, namely, we cannot confirm that branch coverage is a good estimator of mutation coverage in complicated systems.
Fourth, we describe how to adapt mutation testing in order to satisfy industrial time constraints and yet preserve its ability to evaluate the quality of test suite, therefore making it useful in a development environment with small, frequent changes.

\vspace{1em}
\noindent
\textbf{Future Work.} There are several ideas following this study that are worthy of further investigation. In this work, we investigate whether mutation testing reveals more weaknesses in the test suite compared to branch coverage. Given the fact that the ultimate goal of software testing is to reveal as many faults as possible, it is interesting to investigate whether test suites optimized for mutation coverage are more successful in finding faults when compared to test suites optimized for branch coverage. Similarly, the validity of mutant sampling can be investigated deeper by comparing the fault detection capability of test suites optimized for sampled and full-set of mutants. In addition, including more mutation operators to increase the density of mutants in code might in turn increase the efficiency of random sampling. Another prospective research topic is to replicate the results of this study using a minimal set of mutants by detecting the redundant mutants through subsumption analysis. The feasibility of performing subsumption analysis on large industrial software is still in question, and therefore, worth further study.